\documentclass[traditabstract,a4]{aa}  

\usepackage{epsfig}
\usepackage{graphicx}
\usepackage{epstopdf}
\usepackage{color}  
\usepackage{array}   
\usepackage{units}  
\usepackage[breaklinks,colorlinks, citecolor=blue]{hyperref}
\usepackage{natbib}  
\usepackage{multirow}   
\usepackage{amsfonts}
\usepackage{amsmath,amssymb}  
\usepackage[english]{babel}
\usepackage[T1]{fontenc}
\usepackage{longtable,lscape}
\usepackage{footnote}

\usepackage{natbib}
\bibpunct{(}{)}{;}{a}{}{,} 

\usepackage{txfonts}
\usepackage[toc,page]{appendix} 
\usepackage{soul}

\def\ds{\displaystyle}

\def\simlt{\lower.5ex\hbox{$\; \buildrel < \over \sim \;$}}
\def\simgt{\lower.5ex\hbox{$\; \buildrel > \over \sim \;$}}

\newcommand{\dd}{{\mathrm{d}}}

\def\simlt{\lower.5ex\hbox{$\; \buildrel < \over \sim \;$}}
\def\simgt{\lower.5ex\hbox{$\; \buildrel > \over \sim \;$}}

\def\tt{\hbox{{\sc const}}}

\begin{document}


\title{Secondary CMB anisotropies from magnetized halos --I: Power spectra of the Faraday rotation angle and conversion rate}
  
\author{N. Lemarchand\thanks{nadege.lemarchand@ias.u-psud.fr}\inst{\ref{ias}} \and  J. Grain\inst{\ref{ias}} \and G. Hurier\inst{\ref{cefca}} \and F. Lacasa\inst{\ref{unige}} \and A. Fert\'e\inst{\ref{jpl}}}

\institute{
 Institut d'Astrophysique Spatiale, CNRS (UMR8617) and Universit\'{e} Paris-Sud 11, B\^{a}timent 121, 91405 Orsay, France\label{ias}  \and Centro de Estudios de F\'\i{s}ica del Cosmos de Arag\'on (CEFCA), Plaza de San Juan, 1, planta 2, 44001 Teruel, Spain\label{cefca} \and D{\'e}partement de Physique Th{\'e}orique and Center for Astroparticle Physics, Universit{\'e} de Gen{\`e}ve, 24 quai Ernest Ansermet, CH-1211 Geneva, Switzerland\label{unige} \and  Jet Propulsion Laboratory, California Institute of Technology, 4800 Oak Grove Drive, Pasadena, California, USA\label{jpl} 
}

\date{}

\abstract {
Magnetized plasmas within halos of galaxies leave their footprint on the polarized anisotropies of the cosmic microwave background. The two dominant effects for astrophysical halos are Faraday rotation generating rotation of the plane of linear polarization, and Faraday conversion inducing a leakage from linear polarization to circular polarization. We revisit these sources of secondary anisotropies by computing the angular power spectra of the Faraday rotation angle and of the Faraday conversion rate by the large scale structures. To this end, we use the halo model and we pay special attention to the impact of magnetic field projections. Assuming magnetic fields of halos to be uncorrelated, we found a vanishing 2-halo term, and angular power spectra peaking at multipoles $\ell\sim10^4$. The Faraday rotation angle is dominated by the contribution of thermal electrons. For the Faraday conversion rate, we found  that both thermal electrons and relativistic, non-thermal electrons contribute equally in the most optimistic case for the density and Lorentz factor of relativistic electrons, while in more pessimistic cases the thermal electrons give the dominant contribution. Assuming the magnetic field to be independent of the halo mass, the angular power spectra for both effects roughly scale with the amplitude of matter perturbations as $\sim\sigma_8^3$, and with a very mild dependence with the density of cold dark matter. Introducing a dependence of the magnetic field strength with the halo mass leads to an increase of the scaling with the amplitude of matter fluctuations, up to $\sim\sigma_8^{9.5}$ for Faraday rotation and $\sim\sigma_8^{15}$ for Faraday conversion for a magnetic field strength scaling linearly with the halo mass.}


\authorrunning{Lemarchand et al.}
\titlerunning{Secondary CMB anisotropies from magnetized halos (I)}

\maketitle

\section{Introduction}
\label{sec:int}
One of the main challenges in observational cosmology is a complete characterisation of the Cosmic Microwave Background (CMB) polarization anisotropies, targetted by a large amount of on-going, being deployed or planned experiments, either from ground or space-borne missions \citep[see for examples][]{Ade:2018sbj,Suzuki:2018cuy}. In full generality, polarized light (in addition to its total intensity, $I$) is described by its linear component encoded in the two Stokes parameters $Q$ and $U$, and its circular component encoded in the parameter $V$. For CMB anisotropies, there is no source of primordial $V$ in the standard cosmological scenario \citep[see however e.g.][for potential primordial sources]{Giovannini:2009ph}, with upper bounds on its r.m.s. of $\sim1\mu$K at ten degrees \citep{Mainini:2013mja,Nagy:2017csq}. Hence, the CMB polarization field is completely described on the sphere by two Stokes parameters, $Q$ and $U$. In the harmonic domain, this field can be described either by using spin-$(2)$ and spin-$(-2)$ multipolar coefficients, or by using gradient, $E$ and curl, $B$ coefficients. From a physical point of view, the gradient/curl decomposition is more natural as it is directly linked to the cosmological perturbations produced in the primordial Universe. For symmetry reasons, at first order, scalar perturbations can produce $E$-modes only and the $B$-modes part of the polarization field is thus a direct tracer of the primordial gravity waves \citep{1997PhRvD..55.1830Z,Kamionkowski:1996ks}. Though such a picture is partially spoilt by the presence of a secondary contribution generated by the gravitational lensing of the $E$-modes polarization~\citep{1998PhRvD..58b3003Z}, its peculiar angular-scale shape and delensing techniques should allow for a reconstruction of the primordial component.

Lensing of the CMB anisotropies is however not the sole source of {\it cosmological} and {\it astrophysical} $E$-to-$B$ conversion. During the propagation of CMB photons from the last scattering surface to our detectors, the plane of linear polarization could be rotated. Such a rotation could be due to Faraday rotation induced by interactions of CMB photons with background magnetized plasmas, with magnetic fields of either cosmological origin \citep{Kosowsky:1996yc,Kosowsky:2004zh,Campanelli:2004pm,Scoccola:2004ke} or astrophysical origin \citep{Takada:2001bm,Ohno:2002hn,Tashiro:2007mf,Tashiro:2008fc}, or due to interactions with pseudoscalar fields \citep{Carroll:1998zi}. Furthermore, even though primordial circular polarization is not present in the CMB in the standard model of cosmology, secondary circular polarization could be produced by Faraday conversion \citep{1969SvA....13..396S,2003PhLB..554....1C,2015PhRvD..92l3506D} or e.g. by non-linear electrodynamics \citep{Montero-Camacho:2018vgs}.\\

With the significant increase of sensitvity of the forthcoming observatories aiming at an accurate mapping of the CMB polarization on wide ranges of angular scales, clear predictions for such additional secondary anisotropies is of importance for many reasons. 

First, they contain some cosmological and/or astrophysical informations, and could thus be used to probe e.g. parity violation in the Universe \citep{2008PhRvD..78j3516L,Lue:1998mq,Pospelov:2008gg,Yadav:2009eb}, or intrahalo magnetic fields or gas evolution at early epochs \citep{Takada:2001bm,Ohno:2002hn,Tashiro:2007mf,Tashiro:2008fc}. 

Second, such a signal should be known to be correctly taken into account for identifying the primordial component of the $B$-mode from such secondary anisotropies (or at least shown to be subdominant at superdegree scales where primordial $B$-mode is expected to peak above the lensing $B$-mode). 

Thirdly, these secondary anisotropies are of importance for lensing reconstruction using CMB polarized data, shown to be more powerful than starting from temperature data in the case of highly sensitive experiments \citep{Okamoto:2003zw,Hirata:2003ka,Ade:2013gez}. Secondary polarized anisotropies in addition to lensing-induced anisotropies  could indeed mimick contributions from the lensing potential thus biasing its reconstruction from $E$- and $B$-modes. This last point is also of relevance for the delensing, either internal \citep{Seljak:2003pn,Carron:2017vfg}, or based on external tracers of the lensing potential such as the Cosmic Infrared Background \citep{Sigurdson:2005cp,Marian:2007sr,Smith:2010gu,Sherwin:2015baa}. \\

For any possible non-primordial sources of CMB anisotropies, we have first to quantitatively predict the induced CMB anisotropies. Second, one can further investigate the amount of cosmological/astrophysical informations they carry, and finally estimate how they may bias the reconstruction of the primordial $B$-mode and the lensing potential reconstruction. In this article we are interested in magnetized plasmas in halos of galaxies  as a source of secondary polarized anisotropies of the CMB, revisiting and amending first estimates in \citet{Tashiro:2007mf,2003PhLB..554....1C}. Observations with e.g. Faraday rotation measurements from polarized point sources suggest that they are magnetized with a coherence length of the size of the halo scale and a typical strength ranging from $1-10~\mu$G \citep{Kim:1989rt,1998A&A...329..809A,Bonafede:2010xg,Bonafede:2009mq}. This implies that the CMB linear polarisation field is rotated - an effect known as Faraday rotation - and converted to circular polarisation - referred to as Faraday conversion. The goal of the present paper is to give an accurate computation of the angular power spectra of the Faraday rotation angle and Faraday conversion rate, the first mandatory step before estimating its impact on CMB secondary anisotropies. \\

This article is organized as follows. We first briefly describe in Sect. \ref{sec:halos} the propagation of CMB photons through a magnetized plasma. We show that for the specific case of halos, the two dominant effects are Faraday rotation and Faraday conversion. This section is also devoted to a brief presentation of the physics  and the statistics of halos. Second in Sect. \ref{sec:cell}, we present our calculation of the angular power spectra of the Faraday rotation angle and the Faraday conversion rate. This is done using the halo model, and we amend previous analytical calculations giving special attention to the statistics of the projected magnetic fields of halos. Our numerical results are provided in Sect. \ref{sec:results} where we discuss the dependence of the angular power spectra with cosmological parameters. We finally conclude in Sect. \ref{sec:conclu}.

Throughout this article, we use the \citet{2016A&A...596A.107P} (PlanckTTTEEE+SIMlow) best fit parameters, namely $\sigma_8=0.8174$, $\Omega_{CDM}h^2=0.1205$, $\Omega_{b}h^2=0.02225$ and $h=0.6693$.

\section{Physics of halos}
\label{sec:halos}

\subsection{Radiative transfer in a magnetized plasma}
\label{ssec:radiativetransfer}
Propagation of radio and millimeter waves in a magnetized plasma has been studied in \citet{1969SvA....13..396S}, and later reassessed in \citet{1998PASA...15..211K,2005epdm.book.....M,2013MNRAS.430.3320H,2008ApJ...688..695S}. In Eq. (1.5) of \citet{1969SvA....13..396S}, the radiative transfer equation for the four Stokes parameters, $(I,Q,U,V)$, is provided in a specific reference frame in which one of the basis vector in the plane orthogonal to the direction of light propagation is given by the magnetic field projected in that plane. The Stokes parameters $(Q,U,V)$ are however reference-frame dependant, and it is thus important to get this equation in an arbitrary reference frame, for at least two reasons. First, we are interested in the Stokes parameter of the CMB light and there is a priori no reason for the reference frame chosen to measure the Stokes parameter to be specifically aligned with the magnetic fields of the many halos CMB photons pass through. One usually makes use of $(\vec{e}_\theta,\vec{e}_\varphi,\vec{n})$ with $\vec{n}$ pointing along the line-of-sight and $\vec{e}_\theta$, $\vec{e}_\phi$ the unit vectors orthogonal to $\vec{n}$ associated to spherical coordinates, and there is no reason for $\vec{e}_\theta$ to be aligned with the projection of the many magnetic fields. Second, we are here interested in computing the two-point correlation function and there is obviously no reason for the chosen reference frame to coincide at two arbitrary selected directions on the celestial sphere with the specific reference frame used in \citet{1969SvA....13..396S}, which clearly differs from directions to directions on the celestial sphere. 

\begin{table*}
\begin{center}
	\begin{tabular}{l|c|cc|cc}\hline\hline
	& & & & & \\
	& $\dot{\tau}$ & $\dot{\phi}^{I\to P}$ & $\dot{\phi}^{I\to V}$ & $\dot{\alpha}$ & $\dot{\phi}^{P\to V}$ \\ \hline
	Thermal electrons & $n^2_e/(\nu^2T^{3/2}_e)$ & $10^{13} (n_eB_\perp)^2/(\nu^4T^{3/2}_e)$ & $10^6 (n^2_e B_\parallel)/(\nu^3T^{3/2}_e)$& $10^5 (n_eB_\parallel)/(\nu^2)$ & $10^{11} (n_e B_\perp^2)/(\nu^3)$ \\
 	& $\sim10^{-36}$ m$^{-1}$ & $\sim10^{-55}$ m$^{-1}$ & $\sim10^{-46}$ m$^{-1}$ & $\sim10^{-23}$ m$^{-1}$ & $\sim10^{-33}$ m$^{-1}$ \\
	& & & & & \\
	Relativistic electrons & $n^{(r)}_eB_\perp^2/\nu^3$ & $n^{(r)}_eB_\perp^2/\nu^3$ & $ n^{(r)}_e B^{5/2}_n/\nu^{7/2}$ & $ n^{(r)}_eB_\parallel/\nu^2$ & $ n^{(r)}_e B_\perp^2/\nu^3$  \\
 	& $\sim10^{-32}$ m$^{-1}$ & $\sim10^{-32}$ m$^{-1}$ & $\sim10^{-36}$ m$^{-1}$ & $\sim10^{-30}$ m$^{-1}$ & $\sim10^{-31}$ m$^{-1}$ \\ \hline\hline
\end{tabular}
\caption{Scaling of the radiative transfer coefficients for thermal electrons and relativistic electrons with the projection of the magnetic fields along or orthogonal to the line-of-sight, the frequency of photons, and the density and temperature of free electrons \citep[adapetd from][]{1969SvA....13..396S}. For thermal electrons, the numerical constants in front of the reported scalings span a large range of values and we provide their value relative to the one for the parameter $\dot{\tau}$. These constants are all of the same order in the case of relativistic electrons. The corresponding values are obtained for the case of halos with $n_e=10$ m$^{-3}$, $T_e=10^7$ K for thermal electrons, and $n^{(r)}_e=10$ m$^{-3}$ for relativistic electrons. In both cases, the magnetic field is set to $B=3$ $\mu$G, and the frequency to $\nu=30$ GHz.}
\label{tab:coeff}
\end{center}
\end{table*}

The radiative transfer equation is written in an arbitrary reference frame by performing an arbitrary rotation of the basis vectors in the plane orthogonal to the light propagation, or equivalently an arbitrary rotation of the magnetic field projected in such a plane \citep{2009ApJ...703..557H}. We denote here by $\theta_B$ the angle between the magnetic field projected on the plane orthogonal to the line-of-sight and the basis vector $\vec{e}_\theta$. By further introducing the spin-$(\pm2)$ field for linear polarization, $P_{\pm2}=Q\pm iU$, this gives
\begin{equation}
	\frac{\dd}{\dd r}\left(\begin{array}{c}
		I\\
		P_2\\
		P_{-2}\\
		V
	\end{array}\right)=\bigg[\mathbf{M}_{\rm{abs}}+\mathbf{M}_{I\to P}+\mathbf{M}_{P\to P}\bigg]\left(\begin{array}{c}
		I\\
		P_2\\
		P_{-2}\\
		V
	\end{array}\right), \label{eq:sazonov}
\end{equation}
with $r$ labelling the path of light. The three matrices encoding the different contributions to radiative transfer are
\begin{align}
	\mathbf{M}_{\rm{abs}}=\left(\begin{array}{cccc}
		\dot{\tau} & 0 & 0 & 0 \\
		0 & \dot{\tau} & 0 & 0 \\
		0 & 0 & \dot{\tau} & 0 \\
		0 & 0 & 0 &\dot{\tau}
	\end{array}\right), \\
	\mathbf{M}_{I\to P}=\left(\begin{array}{cccc}
		0 & \dot{\phi}^{I\to P}e^{-2i\theta_B} & \dot{\phi}^{I\to P}e^{2i\theta_B} & \dot{\phi}^{I\to V} \\
		\dot{\phi}^{I\to P}e^{2i\theta_B} & 0 & 0 & 0 \\
		\dot{\phi}^{I\to P}e^{-2i\theta_B} & 0 & 0 & 0 \\
		\dot{\phi}^{I\to V} & 0 & 0 &0
	\end{array}\right), \\
	\mathbf{M}_{P\to P}=\left(\begin{array}{cccc}
		0 & 0 & 0 & 0 \\
		0 & 0 & -2i\dot{\alpha} & -i\dot{\phi}^{P\to V}e^{2i\theta_B} \\
		0 & 2i\dot{\alpha} & 0 & i\dot{\phi}^{P\to V}e^{-2i\theta_B} \\
		0 & i\dot{\phi}^{P\to V}e^{-2i\theta_B} & -i\dot{\phi}^{P\to V}e^{2i\theta_B} & 0
	\end{array}\right),
\end{align}
where $\dot{f}$ means differentiation with respect to $r$. 

The different coefficients, $\dot{\tau}$, $\dot{\alpha}$ and the $\dot{\phi}^{i\to j}$'s are real and their expressions can be found in e.g. \citet{1969SvA....13..396S} by setting $\theta_B =0$, which basically corresponds to choosing the specific reference frame adopted in \citet{1969SvA....13..396S}. They are interpreted as follows. First, the coefficient $\dot{\tau}$ in $\mathbf{M}_{\mathrm{abs}}$ simply corresponds to absorption of light by the medium. Second in $\mathbf{M}_{I\to P}$, the coefficients $\dot{\phi}^{I\to P}$ and $\dot{\phi}^{I\to V}$ amount the transfer from total intensity to linear polarization and to circular polarization respectively. Finally in $\mathbf{M}_{P\to P}$, the coefficient $\dot{\alpha}$ corresponds to Faraday rotation which mixes the two modes of linear polarization, while $\dot{\phi}^{P\to V}$ is Faraday conversion which transfers linear polarization in circular polarization. \\

The expressions of the different coefficients and their relative amplitude depend on the nature of free electrons in the magnetized plasma. Two extreme situations are either normal waves of the plasma are circularly polarized, or these normal waves are linearly polarized. In the former case, Faraday rotation is dominant, which is the case for a plasma made of non-relativistic electrons.\footnote{This population is dubbed "cold plasma" in \citet{1969SvA....13..396S}.} In the latter, Faraday conversion is dominant. This can occur for a population of relativistic and non-thermal electrons, with some restrictions on their energy distributions \citep[see][]{1969SvA....13..396S}. 

For the case of astrophysical clusters and halos as considered as magnetized plasmas, two populations of electrons are at play. First, the thermal electrons which are e.g. at the origin of the thermal Sunyaev-Zel'dovich (tSZ) effect, and second, relativistic electrons generated by either AGN or shocks.  For the case of thermal electrons, the typical temperature of clusters is $\sim10^7$ K, corresponding to about few keV's, hence much smaller than the electron mass. This population of electrons is thus mainly non-relativistic. A typical value of the number density of thermal electrons for clusters is $n_e\sim10^3$ m$^{-3}$ for a halo mass of $10^{14}\ensuremath{M_{\odot}}$. For the case of relativistic electrons, the coefficients depend on the energy distribution of the relativistic electrons in the injected plasma via the minimal Lorentz factor, $\Gamma_\mathrm{min}$, and the spectral index of the energy distribution, i.e. $n^{(r)}_e(\Gamma)\propto\Gamma^{~-\beta_E}$, as well as on the spatial distribution of the energy distribution of the injected relativistic electrons in the plasma. We here follow \citet{2003PhLB..554....1C,2015PhRvD..92l3506D} by considering a spectral index of the energy distribution of relativistic electrons of 2, a minimal value of the Lorentz factor of $\Gamma_\mathrm{min}=300$, and an isotropic spatial distribution. The number density of relativistic electrons is largely unknown and we consider here the maximum value we found in the literature, $n^{(r)}_e=10~\mathrm{m}^{-3}$ \citep{Colafrancesco:2002zq}. 

The expressions of these radiative transfer coefficients from \citet{1969SvA....13..396S} are provided in Tab. \ref{tab:coeff} up to numerical constants. We highlight their scaling with the electron number density ($n_e$ and $n^{(r)}_e$), the magnetic field either projected on the line-of-sight, $B_\parallel$, or in the plane perpendicular to it, $B_\perp$, the frequency of the radiation light, $\nu$, and, for the case of thermal electrons the temperature of electrons, $T_e$. For thermal electrons, the numerical constants in front of the reported scalings span a large range of values and we provide their value relative to the one for the parameter $\dot{\tau}$. These constants are all of the same order in the case of relativistic electrons. The values reported are for a magnetic field of $3~\mu\mathrm{G}$ and a frequency of $30~\mathrm{GHz}$. 

For linear polarization, the dominant effect is Faraday rotation by thermal electrons. Faraday rotation from relativistic electrons is 7 orders of magnitude smaller, and absorption, $\dot{\tau}$, is 13 (thermal electrons) and 9 (relativistic electrons) orders of magnitude smaller than Faraday rotation. Faraday conversion from $V$ to $P_{\pm2}$ is zero for CMB since there is no primordial circular polarization. Intensity of the CMB is about 1 to 2 orders of magnitude higher than the $E$-mode of linear polarization, and at least 3 orders of magnitude higher than the $B$-mode. Leakages of $I$ to $P_{\pm2}$ could thus rapidly become important because of this great hierarchy. However, the transfer coefficient $\dot{\phi}^{I\to P}$ for thermal electrons and relativistic electrons is 32 and 9 (resp.) orders of magnitude smaller than $\dot{\alpha}$. Hence leakages from intensity to linear polarization is totally negligible as compared to Faraday rotation by thermal electrons.  

The dominant effect for circular polarization is Faraday conversion from both thermal electrons and relativistic electrons. Absorption is vanishing for zero initial $V$. Leakages from intensity to circular polarization remains smaller than Faraday conversion. In the most optimistic case for the number density of relativistic electrons $\dot{\phi}^{I\to V}$ indeed remains 5 orders of magnitude smaller than $\dot{\phi}^{P\to V}$, meaning that circular polarization generated through leakages of intensity is about 3 orders of magnitude smaller than the one generated through Faraday conversion.\footnote{Note that here Faraday conversion and Faraday rotation by {\it relativistic} electrons are of equal magnitude. Faraday rotation by this population remains however much smaller than the one due to thermal electrons.} \\

An important last comment is in order here. The terms $e^{\pm2i\theta_B}$ naturally appear for preserving the symmetry properties of the four Stokes parameters. 

One reminds that these parameters are defined in the plane $(\vec{e}_{\theta},\vec{e}_{\varphi})$ orthogonal to the line-of-sight and in a reference-frame-dependant manner. The total intensity $I$ is independant of rotation and parity transformations of the reference frame (i.e. it is a {\it scalar}). Linear polarization, $P_{\pm2}$, are {\it spin-${\pm2}$} fields meaning that they rotate by an angle $(\pm2\theta)$ by a rotation $\theta$ of the reference frame, and spin-$(+2)$ and spin-$(-2)$ are interchanged by a parity transformation. Finally, circular polarization $V$ is unchanged through rotations but changes its sign via a parity transformation of the reference frame (i.e. it is a {\it pseudo-scalar}). 

The coefficients $\dot{\alpha}$ and $\dot{\phi}^{P\to V}$ are independent of the reference frame. The angle $\theta_B$ however is reference-frame dependent and the quantities $e^{\pm2i\theta_B}$ are spin-$(\pm2)$ fields. One can then check that indeed all the symmetry properties are properly preserved through radiative transfer. For example, one obtains 
\begin{eqnarray}
	\dot{V}(\vec{n})&=&i\dot{\phi}^{P\to V}(\vec{n})\left[e^{-2i\theta_B(\vec{n})}P_2(\vec{n})-e^{2i\theta_B(\vec{n})}P_{-2}(\vec{n})\right],
\end{eqnarray}
where the right-hand-side is an appropriate combination of different spin-$(\pm2)$ fields leading to a pseudo-scalar field, $V$. We note that this is in agreement with expressions used in \citet{Montero-Camacho:2018vgs,Kamionkowski:2018syl}, reading $\dot{V}(\vec{n})=\phi_U(\vec{n})Q(\vec{n})-\phi_Q(\vec{n})U(\vec{n})$ with $\phi_Q=2\dot{\phi}^{P\to V}\cos(2\theta_B)$ and $\phi_U=2\dot{\phi}^{P\to V}\sin(2\theta_B)$.

It is also easily checked that by selecting the specific reference frame adopted in \citet{1969SvA....13..396S}, i.e. setting $\theta_B=0$, the Eq. (1.5) of  \citet{1969SvA....13..396S} is recovered. In particular in this reference frame one sees that $I$ is transferred into $Q$ only, while $V$ receives contribution from $U$ only, i.e. $\dot{V}=-2\dot{\phi}^{P\to V}U$. (We note that this last expression was used in \citet{2003PhLB..554....1C,2015PhRvD..92l3506D} which is however valid on a very specific reference frame.)

\subsection{Impact on CMB polarization}
The impact of radiative transfer within magnetized halos on the CMB is in theory obtained by integrating Eq. (\ref{eq:sazonov}). Such radiative-transfer distortions of the CMB within halos are expected to mainly occur at low redshifts, $z\lesssim1$. One can thus take as initial conditions the lensed CMB fields.  

In full generality, the matrix $\bigg[\mathbf{M}_{\rm{abs}}+\mathbf{M}_{I\to P}+\mathbf{M}_{P\to P}\bigg]$ is too complicated to find a general solution of this.The dominant effect is however the Faraday rotation by thermal electrons. Neglecting the other coefficients, only linear polarization is modified and the solution is 
\begin{equation}
	P^{\mathrm{FR}}_{\pm2}(\vec{n})=e^{\mp2i\alpha(0,r_\mathrm{CMB})}\,P_{\pm2}(\vec{n}),
\end{equation}
with $P_{\pm2}$ the $\{$primary+lensed$\}$ CMB linear polarization field, and $\alpha(0,r_\mathrm{CMB})$ is the integral of $\dot{\alpha}$ over the line-of-sight from the last scattering surface at $r_\mathrm{CMB}$, to present time at $r=0$ (note that the angle is also a function of $\vec{n}$). The Faraday rotation remains a tiny effect and one can Taylor expand the exponential for small $\alpha$'s.

The next-to-leading order effect is the Faraday conversion whose impact on the CMB can be implemented with a perturbative approach to solve for Eq. (\ref{eq:sazonov}). Since the initial $V$ parameter is vanishing, this leaves the solution for linear polarization unchanged. Circular polarization generated should in principle be generated by Faraday conversion of the rotated linear polarization, $P^{\mathrm{rot}}_{\pm2}$, integrated over the line-of-sight, hence mixing the rotation angle and the conversion rate. These effects are however expected to be small. Multiplicative effect of rotation and conversion are thus of higher orders and it can be neglected. This gives for circular polarization
\begin{equation}
	V(\vec{n})=i\left[\phi_{-2}(0,r_\mathrm{CMB})P_{2}(\vec{n})-\phi_{2}(0,r_\mathrm{CMB})P_{-2}(\vec{n})\right],
\end{equation}
with $\phi_{\pm2}(0,r_\mathrm{CMB})$ the integral over the line-of-sight of $\dot{\phi}^{P\to V}e^{\pm2i\theta_B}$.\footnote{A perturbative approach to solve for Eq. (\ref{eq:sazonov}) keeping $\dot{\alpha}$ at the leading order and $\dot{\phi}^{P\to V}$ at the next-to-leading order gives
\begin{eqnarray}
	P^{\mathrm{FR+FC}}_{\pm2}=e^{\mp2i\alpha(0,r_\mathrm{CMB})}\,P_{\pm2}\mp i\ds\left[\int^0_{r_\mathrm{CMB}}\dd s\dot{\phi}^{P\to V}(s)e^{\pm2i\theta_B(s)}e^{\mp2i\alpha(0,s)}\right]V, \nonumber
\end{eqnarray}
and
\begin{eqnarray}
	V^{\mathrm{FR+FC}}&=&V(\vec{n}) +i\left[\int^0_{r_\mathrm{CMB}}\dd s\dot{\phi}^{P\to V}(s)e^{-2i\theta_B(s)}e^{-2i\alpha(s,r_\mathrm{CMB})}\right]P_2 \nonumber \\
	&-&i\ds\left[\int^0_{r_\mathrm{CMB}}\dd s\dot{\phi}^{P\to V}(s)e^{2i\theta_B(s)}e^{2i\alpha(s,r_\mathrm{CMB})}\right]P_{-2}. \nonumber
\end{eqnarray}
$P_{\pm2}$ and $V$ are the $\{$primary+lensed$\}$ CMB polarization field. Solutions given in the core of the text are obtained setting the initial circular polarization to zero, $V=0$, and keeping the leading order in a Taylor expansion of $e^{\pm2i\alpha(s,r_\mathrm{CMB})}$.}

\subsection{Halos description}
\label{sec:halosphys}
Distortions of the CMB polarized anisotropies by Faraday rotation and Faraday conversion is a multiplicative effect. Their impact on the CMB angular power spectra will thus be determined by the angular power spectra of the Faraday rotation angle, $\alpha$, and the Faraday conversion rate, $\phi_{\pm2}$. 

We make use of the halo model  \citep{Cooray:2002dia} in order to characterize the statistical properties of the radiative transfer coefficients of the halos as magnetized plasmas. The basic elements in this theoretical framework are first the physics internal to each halos, i.e. its gas and magnetic field distributions, and second the statistical properties of halos within our Universe.

\subsubsection{Gas and magnetic field distribution}
In the following, we will mainly need two characteristics of halos: their free electron density and magnetic field spatial profiles, which for simplicity are considered as spherically symmetric.

For the profile $n_e$ of free electrons we choose to take the $\beta$-profile of \citet{1978A&A....70..677C} as what was done in \citet{Tashiro:2007mf}:
\begin{equation}
n_e(r)=n_e^{(c)}\,\left(1+\frac{r^2}{r_c^2}\right)^{-3\beta/2}
\end{equation}
where $r$ and $r_c$ are respectively the physical distance to the halo centre and the typical core radius of the halo (note that it could be comoving distances as only the ratio of these two distances shows up in the expression). The physical halo core radius $r_c$ is related to the virial radius by: $r_{vir}\sim10 r_c$, with $r_{vir}=(M/(4\pi\Delta_c(z)\bar{\rho}(z)/3))^{1/3}$ and $\Delta_c(z)=18\pi^2\Omega_m(z)^{0.427}$ is the spherical overdensity of the virialized halo, $\bar{\rho}(z)$ is the critical density at redshift $z$ \citep[see][]{Tashiro:2007mf}. The quantity $n_e^{(c)}$ is the central free electron density. For thermal free electrons, it is given by: 
\begin{eqnarray}
	n_e^{(c)}&=&9.26\times10^{-4} \ \mathrm{cm}^{-3} \,\left(\frac{M}{10^{14}M_\odot}\right)\,\left(\frac{r_{vir}}{1\mathrm{Mpc}}\right)\,\left(\frac{\Omega_b}{\Omega_m}\right) \\
	&&\times\,{_2F}^{-1}_1(3/2,3\beta/2;5/2;-(r_{vir}/r_c)^2), \nonumber
\label{concentration}
\end{eqnarray}
with $_2F_1$ the hypergeometric function. 

The properties of relativistic electrons inside halos are not well known. Hence we just take a constant value for the central free electron density: $n_e^{(c)}=10$ m$^{-3}$, which was the highest value we found in the literature \citep{Colafrancesco:2002zq}.\\

The magnetic field, denoted $\vec{B}$, is in full generality a function of both $\vec{x}$ and $\vec{x}_i$ (respectively labelling any position within the halo and the center of the halo), as well as a function of the mass and the redshift of the considered halo. Because we have only a poor knowledge of the magnetic field inside halos, we allow ourselves to chose a model for $\vec{B}$ that will simplify a bit the calculations of the angular power spectra. Therefore, the first of our assumptions is that the orientation of the magnetic field is roughly constant over the halo scale, though we still allow for potentially radial profile for its amplitude, i.e. $\vec{B}(\vec{x},\vec{x}_i)=B(\left|\vec{x}-\vec{x}_i\right|)\,\vec{\hat{b}}(\vec{x}_i)$. The vector $\vec{\hat{b}}(\vec{x}_i)$ is a unit vector labelling the orientation of the magnetic field of a given halo, thus depending on the halo position only and considered as a random variable. Here, we also assumed a spherically symmetric profile for the amplitude of the magnetic field. Observations suggest that the amplitude of the magnetic field scales radially as the halo matter content, i.e. $B\propto (n_\mathrm{gas})^\mu$ \citep[see e.g.][]{1991A&A...248...23H,2004A&A...424..429M,Bonafede:2010xg,Bonafede:2009mq}. We thus choose a form for the amplitude of the magnetic field that corresponds to the $\beta$-profile:
\begin{equation}
B(r)=B_c(z)\,\left(1+\frac{r^2}{r_c^2}\right)^{-3\beta\mu/2},
\label{B_profile}
\end{equation}
where $B_c$ is the mean magnetic field strength at the centre of the halo. Its time evolution is given by \citep{2002RvMP...74..775W}:
\begin{equation}
B_c(z)=B_0\,\mathrm{exp}\left(-\frac{t_0-t(z)}{t_d}\right) \mathrm{\,\mu G}
\end{equation}
where $t_0$ is the present time and $t_d=\sqrt{r_{\mathrm{vir}}^3/GM}$, and $B_0$ is the field strength at present time.

\subsubsection{Statistical distribution of halos}

The spatial distribution of halos and their abundance in mass and redshift is described using the halo model \citep{Cooray:2002dia}. The abundance in mass and redshift is given by the halo mass function, $\dd N/\dd M$, and their spatial correlation is derived by the matter power spectrum plus halo bias. In this study, we make use of the halo mass function derived in \citet{2016MNRAS.456.2486D} which is defined using the virial mass. \\

The radiative transfer coefficients introduced in Sect. \ref{ssec:radiativetransfer} depend on the projection of the magnetic field either along the line-of-sight, or in the plane orthogonal to it. One thus needs to introduce some statistics for the orientation of halo's magnetic fields. This statistics of the relative magnetic field orientations of halos is however poorly known. To motivate our choice (presented latter), let us first birefly comment on previous results obtained in the literature. 

The angular power spectrum of the Faraday rotation angle has been firstly computed in \citet{Tashiro:2007mf}, using an approach adapted from the study of the Sunyaev-Zel'dovich effect developed in \citet{1988MNRAS.233..637C,1993ApJ...405....1M,Komatsu:1999ev}. We however believe that this first prediction should be amended. This is motivated by the following intuitive idea (most easily formulated using the 2-point correlation function). 

The Faraday rotation angle is derived from the projection of the magnetic field on the light-of-sight followed by CMB photons, i.e. $\alpha(\vec{n})\propto \vec{n}\cdot\vec{B}$, and the correlation function is thus $\xi(\vec{n}_1,\vec{n}_2):=\left<\alpha(\vec{n}_1)\alpha(\vec{n}_2)\right>\propto \left<\left(\vec{n}_1\cdot\vec{B}_i\right)\left(\vec{n}_2\cdot\vec{B}_j\right)\right>$, where the subscripts $i,~j$ label the halos which are respectively crossed by the line-of-sight $\vec{n}_1$ and $\vec{n}_2$. A first case is that the line-of-sight are such that they cross two distinct halos, i.e. $i\neq j$, corresponding to the so-called 2-halos term in the angular power spectrum. One further assumes that magnetic fields in halos are produced by astrophysical processes. Hence two different halos are statistically independent (from the viewpoint of magnetic fields), leading to $\xi^{\mathrm{2h}}(\vec{n}_1,\vec{n}_2)\propto\left<\vec{n}_1\cdot\vec{B}_i\right>\left<\vec{n}_2\cdot\vec{B}_{j\neq i}\right>$. To be in line with a statistically homogeneous and isotropic Universe, the orientation of the magnetic field of halos should be uniformly distributed leading to $\left<\vec{n}\cdot\vec{B}_{i}\right>=0$.\footnote{Note that for two distinct halos having though the same mass and being at the same redshift, it may well be that they share the same {\it amplitude} for $\vec{B}$. This remains consistent with a statistically homogeneous and isotropic Universe as long as the {\it orientations} of the magnetic fields average down to zero.} One thus expects the 2-halos term to be zero, which is however not the case in \citet{Tashiro:2007mf} where such a term is not vanishing. \footnote{We mention that the 2-halos term may not be vanishing assuming some correlations between the magnetic fields of two different halos (for example if these magnetic fields are seeded by a primordial magnetic field). In this case however, the 2-halos term should be composed of a convolution of the matter power spectrum with the magnetic field power spectrum, as one could expect from results obtained for the similar case of the kinetic Sunyaev-Zel'dovich effect induced by the peculiar velocity of halos \citep{HernandezMonteagudo:2005ys}.}

Considering then the 1-halo term, this reads $\xi^{\mathrm{1h}}(\vec{n}_1,\vec{n}_2)\propto \left<\left(\vec{n}_1\cdot\vec{B}_i\right)\left(\vec{n}_2\cdot\vec{B}_i\right)\right>$ providing that both line-of-sight cross the same halo. This is a priori non-zero since $\left<\vec{B}_i\vec{B}_i\right>$ does not vanish. There is however a subtlety which to our viewpoint, has not been considered in \citet{Tashiro:2007mf}. They consider that the statistical average of the orientation of magnetic fields for the 1-halo term is $\left<\left(\vec{n}\cdot\vec{B}_i\right)^2\right>=1/3$ (the value being the one corresponding to uniformly distributed orientations). However, the spatial extension of halos allows for having two {\it different} lines-of-sight crossing the {\it same} halo, and there is a priori no reason that $\left(\vec{n}_1\cdot\vec{B}_i\right)=\left(\vec{n}_2\cdot\vec{B}_i\right)$ for a randomly selected halo. As a consequence, this is $\left<\left(\vec{n}_1\cdot\vec{B}_i\right)\left(\vec{n}_2\cdot\vec{B}_i\right)\right>$ which enters as a statistical average on the1-halo term, and not $\left<\left(\vec{n}\cdot\vec{B}_i\right)^2\right>$. 

A similar argument applies for Faraday conversion except that this the projection of the magnetic field on the plane orthogonal to $\vec{n}$ which is here involved.\\

We will thus suppose that orientations are uniformly distributed in the Universe, independant for two different halos, and independant of the spatial distribution of halos. This can be understood as follows: we assume no coherence of the magnetic field orientations of different halos or, to put it differently, the magnetic field correlation length is smaller than the inter-halo scale. This assumption is clearly in line with the cosmological principle, and it is motivated by the idea that halos' magnetism is a result of processes isolated from other halos. Thus, this orientation is a random variable which should be zero once averaged over halos. 

Orientations are given by the unit vector, $\vec{b}$, which is thus labelled by a zenithal angle, $\beta(\vec{x}_i)$, and an azimuthal angle, $\alpha(\vec{x}_i)$. In the cartesian coordinate system, the three component are
\begin{eqnarray}
 b_x^{i}&=&\sin\left(\beta(\vec{x}_i)\right)\cos\left(\alpha(\vec{x}_i)\right),\\
 b_y^{i}&=&\sin\left(\beta(\vec{x}_i)\right)\sin\left(\alpha(\vec{x}_i)\right),\\
 b_z^{i}&=&\cos\left(\beta(\vec{x}_i)\right).
\end{eqnarray}
Any projection of the magnetic field orientation can be written as a function of the two angles, $\beta$ and $\alpha$. Our assumption of uniformly distributed orientations translates into the following averaging 
\begin{eqnarray}
	\left<f(\alpha^i,\beta^i)\right>&=&\frac{1}{4\pi}\displaystyle\int f(\alpha^i,\beta^i)\,\dd\alpha^i\,\dd(\cos\beta^i),
\end{eqnarray}
with $\beta^i$ and $\alpha^i$ a shorthand notation for $\beta(\vec{x}_i)$ and $\alpha(\vec{x}^i)$. Since we assume two halos to be independant, one does not need to further introduce some correlations and the above fully described the statistics of orientations of magnetic fields.

\section{Angular power spectra of Faraday rotation and Faraday conversion}
\label{sec:cell}
\subsection{Faraday rotation angle}
The Faraday rotation angle is given by the following integral over the line-of-sight
\begin{equation}
	\alpha(\vec{n})=\frac{e^3}{8\pi^2\,m_e^2\,c\,\varepsilon_0}\ds\int_0^{r_{\mathrm{CMB}}}\frac{a(r)\dd r}{\nu^2(r)}\,\sum_{i=\mathrm{halo}}\left[\vec{\hat{n}}\cdot\vec{B}(\vec{x},\vec{x}_i)\right]\,n_e(\left|\vec{x}-\vec{x}_i\right|),
\end{equation}
where $r$ stands for the comoving distance on the line-of-sight, $\vec{x}=r\vec{n}$, $r_{\mathrm{CMB}}$ is the distance to the last-scattering surface, and $\vec{x}_i$ is the center of the $i$-th halo. With our assumption regarding the magnetic field, and further replacing the summation over halos by integrals over the volume and over the mass range, the above reads
\begin{eqnarray}
	\alpha(\vec{n})=\frac{e^3}{8\pi^2\,m_e^2\,c\,\varepsilon_0}\ds\int_0^{r_{\mathrm{CMB}}}\frac{a(r)\dd r}{\nu^2(r)}\,\iint \dd M_i \dd^3\vec{x}_i ~\Big[n_h(\vec{x}_i) \label{eq:faradaydef} \\
	  \times\, b(\vec{n},\vec{x}_i)X\left(\left|\vec{x}-\vec{x}_i\right|\right)\Big], \nonumber
\end{eqnarray}
with $n_h(\vec{x}_i)$ the abundance of halos, $b(\vec{n},\vec{x}_i)=\vec{n}\cdot\vec{b}(\vec{x}_i)$ the projection along the line-of-sight, and $X\left(\left|\vec{x}-\vec{x}_i\right|\right)=B\left(\left|\vec{x}-\vec{x}_i\right|\right)n_e\left(\left|\vec{x}-\vec{x}_i\right|\right)$. \\

Two simplifications result from the different assumptions made about the statistics of the orientation of the magnetic field. To this end, let us introduce the notation
$$
	A^i(\vec{n})=\frac{e^3}{8\pi^2\,m_e^2\,c\,\varepsilon_0}\ds\int_0^{r_{\mathrm{CMB}}}\frac{a(r)\dd r}{\nu^2(r)}\,X\left(\left|\vec{x}-\vec{x}_i\right|\right),
$$
where we stress that the impact of orientation is omitted in the above. It can basically be interpreted as the maximum amount of rotation the halo $i$ can generate. (We note that this is also a function of the mass of the halo.)

	In the halo model first, the angular power spectrum, or equivalently the 2-point correlation function, is composed of a 1-halo term and a 2-halo term. This gives for the 1-halo term
\begin{eqnarray}
	\left<\alpha(\vec{n}_1)\alpha(\vec{n}_2\right>_{1\mathrm{h}}&=&\ds\iint\dd M_i\dd^3\vec{x}_i \left(\frac{\dd N}{\dd M}\right)A^i(\vec{n}_1)A^i(\vec{n}_2) \\
	&&\times\left<b(\vec{n}_1,\vec{x}_i)b(\vec{n}_2,\vec{x}_i)\right>, \nonumber
\end{eqnarray}
where we use $\left<n^2_h(\vec{x}_i)\right>=\dd N/\dd M$.\footnote{We remind that abundances are given by a Poisson staistics for which $\left<n^2_h\right>=\left<n_h\right>$.} The 2-halo term then reads
\begin{eqnarray}
	\left<\alpha(\vec{n}_1)\alpha(\vec{n}_2)\right>_{2\mathrm{h}}&=&\ds\iint\dd M_i\dd^3\vec{x}_i\iint\dd M_j\dd^3\vec{x}_j\left<n_h(\vec{x}_i)n_h(\vec{x}_j)\right> \nonumber \\
	&&\times A^i(\vec{n}_1)A^j(\vec{n}_2)\left<b(\vec{n}_1,\vec{x}_i)b(\vec{n}_2,\vec{x}_j)\right>,
\end{eqnarray}
where in the above the halo $j$ is necessarily different from the halo $i$.\footnote{Note that in the above $$\left<n_h(\vec{x}_i)n_h(\vec{x}_j)\right>=\left(\frac{\dd N}{\dd M_i}\right)\left(\frac{\dd N}{\dd M_j}\right)\left[1+b(M_i,z_i)b(M_j,z_j)\xi_\mathrm{m}(\vec{x}_i-\vec{x}_j\right),$$ with $b(M,z)$ the bias and $\xi_\mathrm{m}$ the 2-point correlation function of the matter density field.} The 2-halo term is however vanishing because of averaging over the orientation of magnetic field. Since two different halos have uncorrelated magnetic fields, one has $\left<b(\vec{n}_1,\vec{x}_i)b(\vec{n}_2,\vec{x}_j)\right>=\Big<b(\vec{n}_1,\vec{x}_i)\Big>\left<b(\vec{n}_2,\vec{x}_j)\right>$, which is finally equal to zero since magnetic orientations have a vanishing ensemble average.

Second, the 2-point correlation function is described by an angular power spectrum, i.e.
\begin{equation}
	\left<\alpha(\vec{n}_1)\alpha(\vec{n}_2)\right>=\ds\sum_\ell C^\alpha_\ell\,Y_{\ell m}(\vec{n}_1)\,Y_{\ell m}^\star(\vec{n}_2).
\end{equation}
As detailed in App. \ref{app:wigner}, this angular power spectrum, $C^\alpha_\ell$, is given by the convolution of two angular power spectra reading
\begin{eqnarray}
	C^\alpha_\ell&=&\frac{1}{4\pi}\ds\sum_{L,L'}\left(2L+1\right)\left(2L'+1\right)\left(\begin{array}{ccc}
		L & L' & \ell \\
		0 & 0 & 0
	\end{array}\right)^2\,D^A_L\,D^\parallel_{L'} \label{eq:frgen}
\end{eqnarray}
where $D^A_L$ is the angular power spectrum associated to the 2-point functions of the maximum of the rotation angle, i.e. $$\ds\iint\dd M_i\dd^3\vec{x}_i \left(\frac{\dd N}{\dd M}\right)A^i(\vec{n}_1)A^i(\vec{n}_2),$$ and $D^\parallel_{L'}$ is the angular power spectrum associated to the correlation function of orientations, $\left<b(\vec{n}_1,\vec{x}_i)b(\vec{n}_2,\vec{x}_i)\right>$. Finally, the term $$\left(\begin{array}{ccc}
		L & L' & \ell \\
		0 & 0 & 0
	\end{array}\right)$$ corresponds to Wigner-$3j$'s. The expression in Eq. (\ref{eq:frgen}) means that the total angular power spectrum is obtained as the angular power spectrum for the maximum amount of the effect, $D^A_L$, modulated by the impact of projecting the magnetic field on the line-of-sight, hence the convolution with $D^\parallel_{L'}$. \\

It is shown in App. \ref{app:dl} that the angular power spectrum $D^A_L$ reads using Limber's approximation
\begin{eqnarray}
	D^A_L&=&\ds\int_0^{z_{\mathrm{CMB}}}\dd z\left(\frac{r}{\nu^2(r)}\right)^2\frac{\dd r}{\dd z}\int \dd M \,\frac{\dd N}{\dd M} \left[\alpha_{(c)}\alpha_L\right]^2, \label{eq:limberdl}
\end{eqnarray}
with $\alpha_{c}(M,z)$ the rotation angle at the core of the halo given by
\begin{equation}
	\alpha_{c}=\left(\frac{e^3}{m^2_ec\varepsilon_0\sqrt{8\pi}}\right)~n^{(c)}_e(M,z)\,B_c(B_0,z).
\end{equation}
This core angle depends on the mass, the redshift and the magnetic field amplitude of the considered halos. The projected Fourier transform of the profile is
\begin{equation}
	\alpha_\ell=\sqrt{\frac{2}{\pi}}\,\left(\frac{r^{\mathrm{(phys)}}_c}{\ell_c^2}\right)\int_0^\infty\dd x \, x^2U(x)\,j_0((\ell+1/2)x/\ell_c),
\end{equation}
with $\ell_c=D_\mathrm{ang}(z)/r_c$ the characteristic multipole for a halo of size $r_c$ at a redshift $z$, and $D_\mathrm{ang}(z)$ the angular diameter distance. The normalized profile $ U(x)$ for a $\beta$-profile is $U(x)=(1+x)^{-3\beta(1+\mu)/2}$ where $x=r/r_c$.

Similarly in App. \ref{app:ofr}, the angular power spectrum for the orientation of the magnetic field projected on the line-of-sight is
\begin{eqnarray}
	D^\parallel_{L'}=\frac{4\pi}{9}\,\delta_{L',1}.
\end{eqnarray}
Using the triangular conditions for the Wigner-$3j$ \citep[see e.g][]{varshalovich1988quantum}, the angular power spectrum of the Faraday rotation angle boils down to
\begin{equation}
	C^\alpha_\ell=\frac{1}{3}\left[\left(\frac{\ell}{2\ell+1}\right) D^A_{\ell-1} +\left(\frac{\ell+1}{2\ell+1}\right)D^A_{\ell+1}\right]. \label{eq:cellfr}
\end{equation}
We note that the above does not assume Limber's approximation in the sense that the involved $D^A_\ell$'s can be either the one obtained from the Limber's approximation, Eq. (\ref{eq:limberdl}), or the non-approximated one as given in Eq. (\ref{eq:fulldl}). \\

The impact of projecting the magnetic fields on the line-of-sight translates into the modulation of the angular power spectra for the maximum amount of rotations halos can generate. In the limit of high values of $\ell$, the two line-of-sights, $\vec{n}_1$ and $\vec{n}_2$, can be considered as very closed one to each other. This leads to $\left<b(\vec{n}_1,\vec{x}_i)b(\vec{n}_2,\vec{x}_i)\right>\simeq\left<b^2(\vec{n}_1,\vec{x}_i)\right>=1/3$ and one should recover the same result as derived in \citet{Tashiro:2007mf}, {\it restricted} to the 1-halo term however. In this high-$\ell$ limit, Eq. (\ref{eq:cellfr}) simplifies to $C^\alpha_\ell=D^A_\ell/3$. From the expression of $D^A_\ell$ using Limber's approximation, one can check that this is identical to the 1-halo term derived in \citet{Tashiro:2007mf}.

\subsection{Faraday conversion}
For the Faraday conversion, one first reminds that irrespectively of the nature of free electrons (either from a thermal distribution or from a relativistic, non-thermal distribution) the conversion rate is proportional to $B^2_\perp e^{\pm2i\theta_B}$ with $B_\perp$ the norm of the projected magnetic field on the plane orthogonal to $\vec{n}$, and $\theta_B$ is the angle between the projected magnetic field and the first basis vector in the plane orthogonal to $\vec{n}$. This defines the spin-$(\pm2)$ structure of these conversion coefficients which can be conveniently rewritten using projections of the magnetic field on the so-called helicity basis in the plane orthogonal to $\vec{n}$, i.e.
\begin{equation}
	B^2_\perp e^{\pm2i\theta_B}=B^2\left(\left|\vec{x}-\vec{x}_i\right|\right)\,\left[\vec{b}(\vec{x}_i)\cdot\left(\vec{e}_\theta\pm i\vec{e}_\varphi\right)\right]^2,
\end{equation}
where we remind in the above that the norm of the magnetic field is a radial function and its orientation depends on the halos location only. 

\subsubsection{Thermal electrons}
The radiative transfer coefficients integrated over the line-of-sight is defined as $\phi_{\pm2}(\vec{n})=\int a(r)\dd r\sum_\mathrm{halos}\dot{\phi}^{P\to V}_i(\vec{n},r)e^{\pm2i\theta_B^{(i)}(\vec{n},r)}$. For thermal electrons, this explicitly reads
\begin{eqnarray}
	\phi_{\pm2}(\vec{n})=\frac{e^4}{16\pi^3m^3_e\,c\,\varepsilon_0}\ds\int_0^{r_{\mathrm{CMB}}} \frac{a(r)}{\nu^3(r)}\dd r\iint\dd M_i\dd\vec{x}_i~\Big[n_h(\vec{x}_i) \\
	\times\,b_{\pm2}(\vec{n},\vec{x}_i)X\left(\left|\vec{x}-\vec{x}_i\right|\right)\Big], \nonumber
\end{eqnarray}
where now $X\left(\left|\vec{x}-\vec{x}_i\right|\right)=n_e\left(\left|\vec{x}-\vec{x}_i\right|\right)\, B^2\left(\left|\vec{x}-\vec{x}_i\right|\right)$, and $b_{\pm2}(\vec{n},\vec{x}_i)=\left[\vec{b}(\vec{x}_i)\cdot\left(\vec{e}_\theta\pm i\vec{e}_\varphi\right)\right]^2$. 

Apart from the spin-$(\pm2)$ structure encoded in $b_{\pm2}$, the above has exactly the same structure as the Faraday rotation angle, Eq. (\ref{eq:faradaydef}), and we adopt the same strategy as for the Faraday rotation angle. The key difference for Faraday conversion lies in the spin structure and one has to compute three correlations (2 autocorrelations and 1 cross-correlation). One can either use spin fields or more conveniently, $E$ and $B$ decompositions which is reference frame independent \citep[see e.g.][]{Kamionkowski:1996ks,1997PhRvD..55.1830Z}. Here we will first compute the correlation for spin fields, defined as 
\begin{eqnarray}
	&&\left<\phi_{\pm2,\ell m}\phi_{\pm2,\ell'm'}^\star\right>=C^{\pm2,\pm2}_\ell\, \delta_{\ell,\ell'}\delta_{m,m'}, \\
	&&\left<\phi_{2,\ell m}\phi_{-2,\ell'm'}^\star\right>=C^{2,-2}_\ell\, \delta_{\ell,\ell'}\delta_{m,m'}.
\end{eqnarray}
These angular power spectra are easily transformed into angular power spectra for the $E$ and $B$  field associated to $\phi_{\pm2}$ using $\phi^E_{\ell m}=-(\phi_{2,\ell m}+\phi_{-2,\ell m})/2$ and $\phi^B_{\ell m}=i(\phi_{2,\ell m}-\phi_{-2,\ell m})/2$. \\

As the case for Faraday rotation, the 2-halo term is vanishing because of the orientations of the magnetic fields averages down to zero, i.e. $$\left<b_{\pm2}(\vec{n}_1,\vec{x}_i)b_{\pm2}(\vec{n}_2,\vec{x}_{j\neq i})\right>=\bigg<b_{\pm2}(\vec{n}_1,\vec{x}_i)\bigg>\,\left<b_{\pm2}(\vec{n}_2,\vec{x}_{j\neq i})\right>$$ for two different halos, and for uniformly random orientations one found that $\left<b_{\pm2}(\vec{n},\vec{x}_i)\right>=0$. 

Following App. \ref{app:wigner} then, one shows that
\begin{equation}
	C^{\pm2,\pm2}_{\ell}=\frac{1}{4\pi}\,\displaystyle\sum_{L,L'}\,(2L+1) (2L'+1)\, \left(\begin{array}{ccc}
		L' & L & \ell \\
		\mp2 & 0 & \pm 2
		\end{array}\right)^2
        ~D^\Phi_L \ D^\perp_{L'},
\end{equation}
and
\begin{eqnarray}
	C^{2,-2}_\ell=\frac{1}{4\pi}\,\displaystyle\sum_{L,L'}\,(2L+1) (2L'+1)~D^\Phi_L \ D^\perp_{L'}, \nonumber \\
	\times\,\left(\begin{array}{ccc}
		L' & L & \ell \\
		-2 & 0 &  2
		\end{array}\right)\left(\begin{array}{ccc}
		L' & L & \ell_1 \\
		2 & 0 &  -2
		\end{array}\right).
\end{eqnarray}
The above is interpreted in a very similar way to $C^\alpha_\ell$. It is the power spectrum of the maximum of the effect of Faraday conversion, $D^\Phi_L$, which is further modulated by the impact of projecting the magnetic field in the plane orthogonal to the line-of-sight, which is encoded in $D^\perp_{L'}$.

The angular power spectrum of the amplitude of the effect is derived using the standard technique reminded in App. \ref{app:dl} and by selecting the appropriate profile, $n_EB^2$ instead of $n_EB$. This gives with the Limber's approximation
\begin{equation}
	D^\Phi_L=\ds\int_0^{z_{\mathrm{CMB}}}\dd z\left(\frac{r}{\nu^3(r)}\right)^2\frac{\dd r}{\dd z}\int \dd M \,\frac{\dd N}{\dd M} \left[\Phi_{(c)}\phi_L\right]^2, \label{eq:dlfc}
\end{equation}
with the amplitude of the conversion at the core of the halo given by
\begin{equation}
	\Phi_{(c)}=\left(\frac{e^4}{2(2\pi)^{3/2}m^3_e\,c\,\varepsilon_0}\right)\, n_e^{(c)}\,B^2_{c}.
\end{equation}
The Fourier-transformed normalized profile reads
\begin{equation}
	\phi_\ell=\sqrt{\frac{2}{\pi}}\,\left(\frac{r^{\mathrm{(phys)}}_c}{\ell_c^2}\right)\int_0^\infty\dd x \, x^2U(x)\,j_0((\ell+1/2)x/\ell_c),
\end{equation}
where the profile is now given by $U(x)=(1+x)^{-3\beta(1+2\mu)/2}$. The angular power spectrum for the orientation contribution is detailed in App. \ref{app:ofc}. It is nonzero for a multipole of 2 only and it reads $D^\perp_{L'}=(32\pi/75)\,\delta_{L',2}$.

The last step consists in deriving the angular power spectrum in the $E$ and $B$ decomposition of the spin-$(\pm2)$ of the Faraday conversion coefficients. This first shows that the $\left<EB\right>$ cross-spectrum is vanishing, i.e. $C^{\phi^E\phi^B}_\ell=0$. The autospectra are given by
\begin{eqnarray}
	C^{\phi^E\phi^{E}}_{\ell}=&\ds\frac{4}{15}&\left[\frac{(\ell+1)(\ell+2)}{2(2\ell-1)(2\ell+1)}\,D^\Phi_{\ell-2}+\frac{3(\ell-1)(\ell+2)}{(2\ell-1)(2\ell+3)}\,D^\Phi_{\ell}\right. \nonumber \\
	&&\left.+\frac{\ell(\ell-1)}{2(2\ell+1)(2\ell+3)}\,D^\Phi_{\ell+2}\right], \label{eq:cellphiE}
\end{eqnarray}
and
\begin{eqnarray}
	C^{\phi^B\phi^{B}}_{\ell}=\frac{4}{15}\left[\left(\frac{\ell+2}{2\ell+1}\right)D^\Phi_{\ell_1-1}+\left(\frac{\ell-1}{2\ell+1}\right)D^\Phi_{\ell+1}\right]. \label{eq:cellphiB}
\end{eqnarray}
In the above, we made use of the triangular conditions for the Wigner-$3j$'s. (We note that the above angular power spectra are spin-$(\pm2)$ and they are nonvanishing for $\ell\geq2$.) In the high-$\ell$ limit, the two autospectra are identical and equal to $C^{\phi_E\phi_E}_\ell\simeq C^{\phi_B\phi_B}_\ell\simeq(4/15)D^\Phi_\ell$.

\subsubsection{Relativistic electrons}
For relativistic electrons, the rate of Faraday conversion integrated over the line-of-sight reads
\begin{eqnarray}
	\phi_{\pm2}(\vec{n})&=&\frac{e^4\Gamma_{\mathrm{min}}}{8\pi^3m^3_e\,c\,\varepsilon_0}\left(\frac{\beta_E-1}{\beta_E-2}\right)\ds\int_0^{r_{\mathrm{CMB}}} \frac{a(r)}{\nu^3(r)}\dd r \\
	&&~~~~~~\iint\dd M_i\dd\vec{x}_i~\Big[n_h(\vec{x}_i)\,b_{\pm2}(\vec{n},\vec{x}_i)X\left(\left|\vec{x}-\vec{x}_i\right|\right)\Big], \nonumber
\end{eqnarray}
where $\Gamma_{\mathrm{min}}$ is the minimum Lorentz factor of the relativistic electrons, and $\beta_E$ is the spectral index of the energy distribution of relativistic electrons. The profile is $X=n_e^{(r)}\,B^2$, i.e. the same as for thermal electrons replacing the number density of thermal electrons by the number density of relativistic ones. 

The angular power spectrum for the Faraday conversion rate due to relativistic electrons has exactly the same form as for thermal electrons, i.e. Eqs. (\ref{eq:cellphiE}) \& (\ref{eq:cellphiB}) for the $E$ and $B$ autospectra. The expression for $D^\Phi_\ell$ also reads the same. It is given by Eq. (\ref{eq:dlfc}) where one just has to replace $\Phi_{(c)}$ by 
\begin{equation}
	\Phi_{(r)}=\left(\frac{e^4\Gamma_{\mathrm{min}}}{4(2\pi)^{3/2}m^3_e\,c\,\varepsilon_0}\right)\left(\frac{\beta_E-1}{\beta_E-2}\right)\, n_e^{(r)}\,B^2_{c}.
\end{equation}

\subsection{Remarks on cross-correlation}
Let us briefly comment on possible cross-correlation. The first point is that in this approach, the cross-correlation between the Faraday rotation angle with any tracer of halos which is not correlated with the projection of magnetic fields on the line-of-sight will be vanishing. This is because the cross-correlation is proportional to either $\left<\vec{b}\cdot\vec{n}\right>$ or $\left<\left[\vec{b}\cdot\left(\vec{e}_\theta\pm i\vec{e}_\varphi\right)\right]^2\right>$, both of which average down to zero. This will be indeed the case for cross-correlation with the thermal and relativistic Sunyaev-Zel'dovich effect, the lensing potential, or the CIB fluctuations. This will also be the case for cross-correlation with the absorption coefficients, $\mu$. 

Finally, we checked that the averages $\left<\left[\vec{b}\cdot\vec{n}_1\right]\left[\vec{b}\cdot\left(\vec{e}^{(2)}_\theta\pm i\vec{e}^{(2)}_\varphi\right)\right]^2\right>$ equals to zero. This yield a vanishing cross-correlation between the Faraday rotation angle and Faraday conversion.

\section{Numerical results}
\label{sec:results}
Our results are shown for a frequency of observation $\nu_0=30$ GHz, and a field strength at present time $B_0=3$ $\mu$G. We remind that the angular power spectrum for the Faraday rotation angle  scales as $C^\alpha_\ell\propto B^2_0/\nu_0^4$. The angular autospectra of the $E$ and $B$ modes of Faraday conversion scales as $C^{\phi^E\phi^E(\phi^B\phi^B)}_\ell\propto B^4_0/\nu_0^6$. 

Unless specified, the parameters for the $\beta$-profile are $\beta=\mu=2/3$, which would correspond to a magnetic field frozen into matter.

All the numerical results reported here are obtained using the universal mass function from \citet{2016MNRAS.456.2486D}. For consistency, we checked that similar results are obtained using the mass function of \citet{2008ApJ...688..709T}. In particular, we found similar scaling with cosmological parameters, despite small variation regarding the overall amplitude of the angular power spectra.

\begin{figure*}
\begin{center}
\includegraphics[scale=0.39]{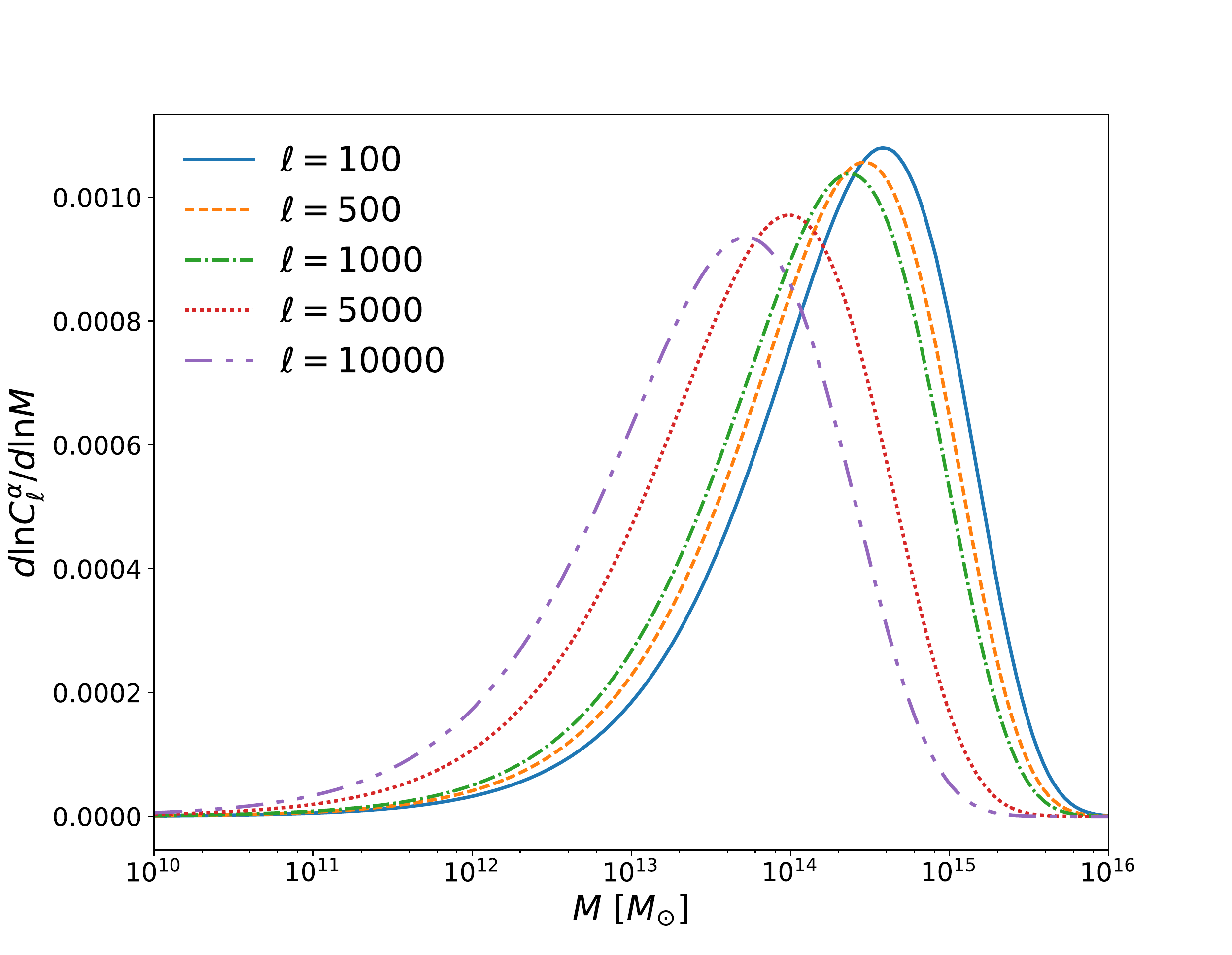}
\includegraphics[scale=0.39]{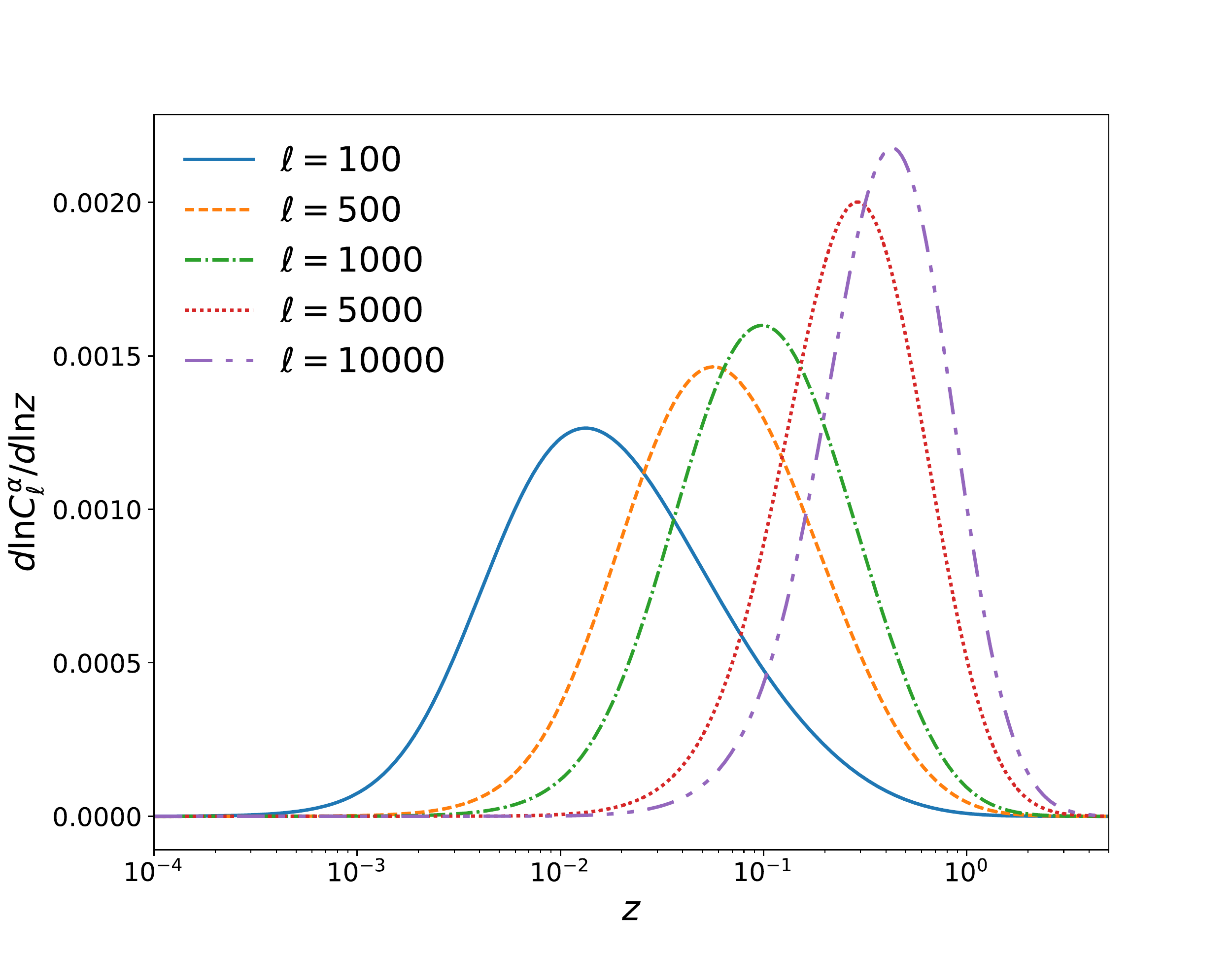}
\caption{\label{mass_redshift_dependence} Left: the mass distribution of the Faraday rotation effect for various $\ell$ modes. Right: the redshift distribution of the Faraday rotation effect for various $\ell$ modes.}
\end{center}
\end{figure*}

\subsection{Power spectrum of the Faraday rotation angle}

Fig. ~\ref{mass_redshift_dependence} shows the mass and redshift distributions of the Faraday rotation angle power spectrum for different multipoles $\ell$, with on the left $\dd\mathrm{ln}C_\ell^\alpha/\dd\mathrm{ln}M$ as a function of mass, and on the right $\dd\mathrm{ln}C_\ell^\alpha/\dd\mathrm{ln}z$ as a function of redshift. Compared to \citet{Tashiro:2007mf} (Fig.~(4) and (3) respectively), we note that our distributions are slightly shifted to higher masses and lower redshifts. This results in the Faraday rotation effect being more sensitive to higher mass values and lower redshift galaxy halos than their Faraday rotation angle, so that its power spectrum seems to be slightly shifted to lower $\ell$ values as compared to the one in \citet{Tashiro:2007mf}: indeed, low multipoles correspond to high angular scales, hence to high masses or low redshifts halos because these halos appear bigger on the sky than low masses and high redshifts halos. 

\begin{figure}
\begin{center}
\includegraphics[scale=0.4]{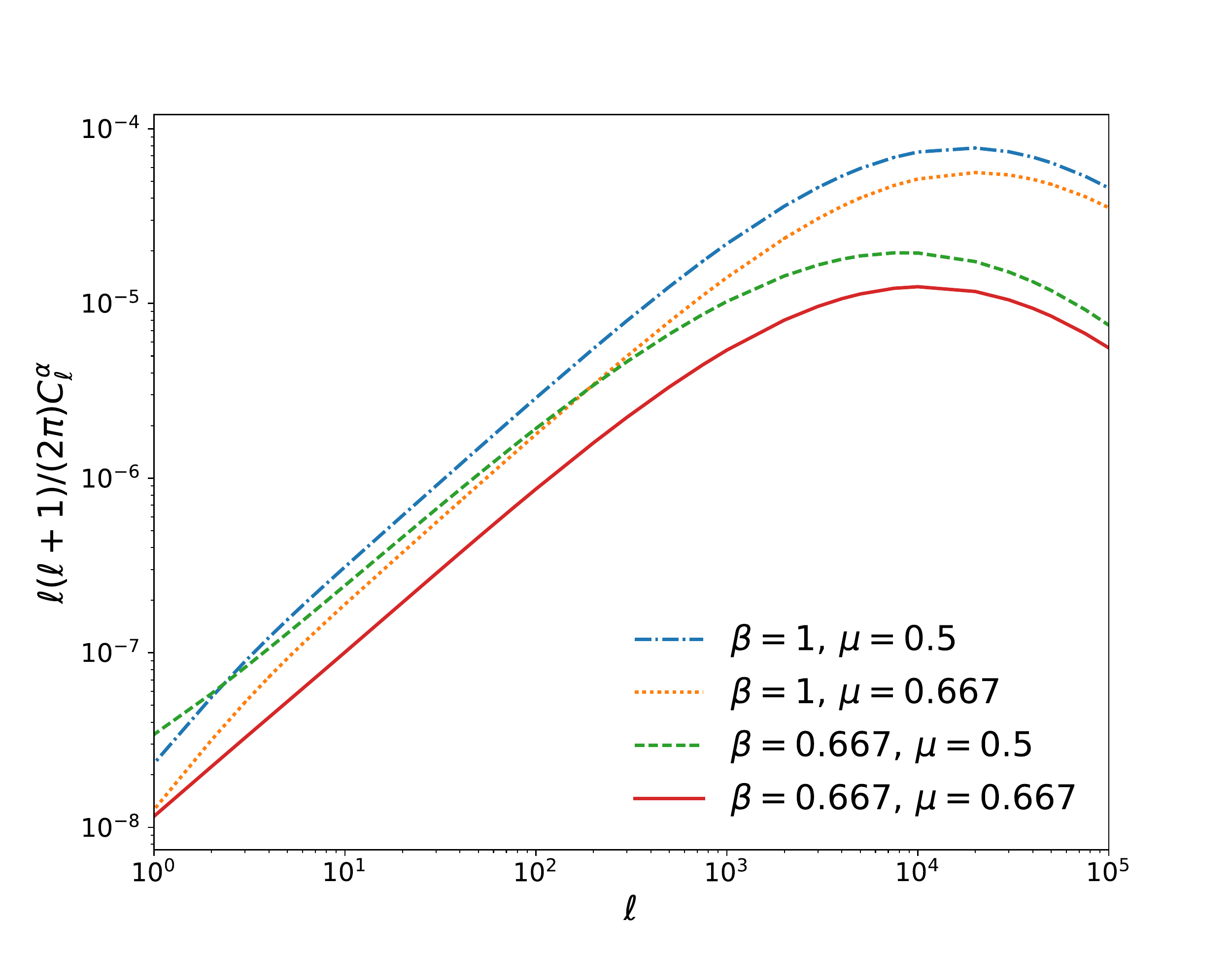}
\caption{\label{angularpowerspectrum_beta-mu} The angular power spectra of the Faraday rotation angle, $C^\alpha_\ell$, for different values of the parameter $\beta$ of the $\beta$-profile and different values of the parameter $\mu$ of the magnetic field profile. }
\end{center}	
\end{figure}

Fig.~\ref{angularpowerspectrum_beta-mu} shows the angular power spectrum of the Faraday rotation angle for different values of the parameters $\beta$ and $\mu$ of the spatial distribution profiles of the free electrons density and magnetic field, respectively. First, one can note a shift of power to higher multipoles when increasing $\beta$ or $\mu$: indeed the profile of free electrons and magnetic fields then becomes steeper so that the they are more concentrated in the centre of the halo which consequently appears smaller on the sky. This result is consistent with \citet{Tashiro:2007mf}. We also see that the difference in amplitudes is more significant when we change $\beta$ rather than $\mu$, because $\beta$ appears both in the free electrons and magnetic field profiles. However, the trend is different when changing $\beta$ or $\mu$. Indeed, when increasing $\mu$, the amplitude decreases, as expected from the magnetic field profile Eq.~(\ref{B_profile}). On the contrary, when increasing $\beta$, the amplitude also increases: this is because as the profile of free electrons is steeper, keeping the number of electrons constant, their concentration increases Eq.~(\ref{concentration}), so does the amplitude. \\

\begin{figure*}
\begin{center}
\includegraphics[scale=0.4]{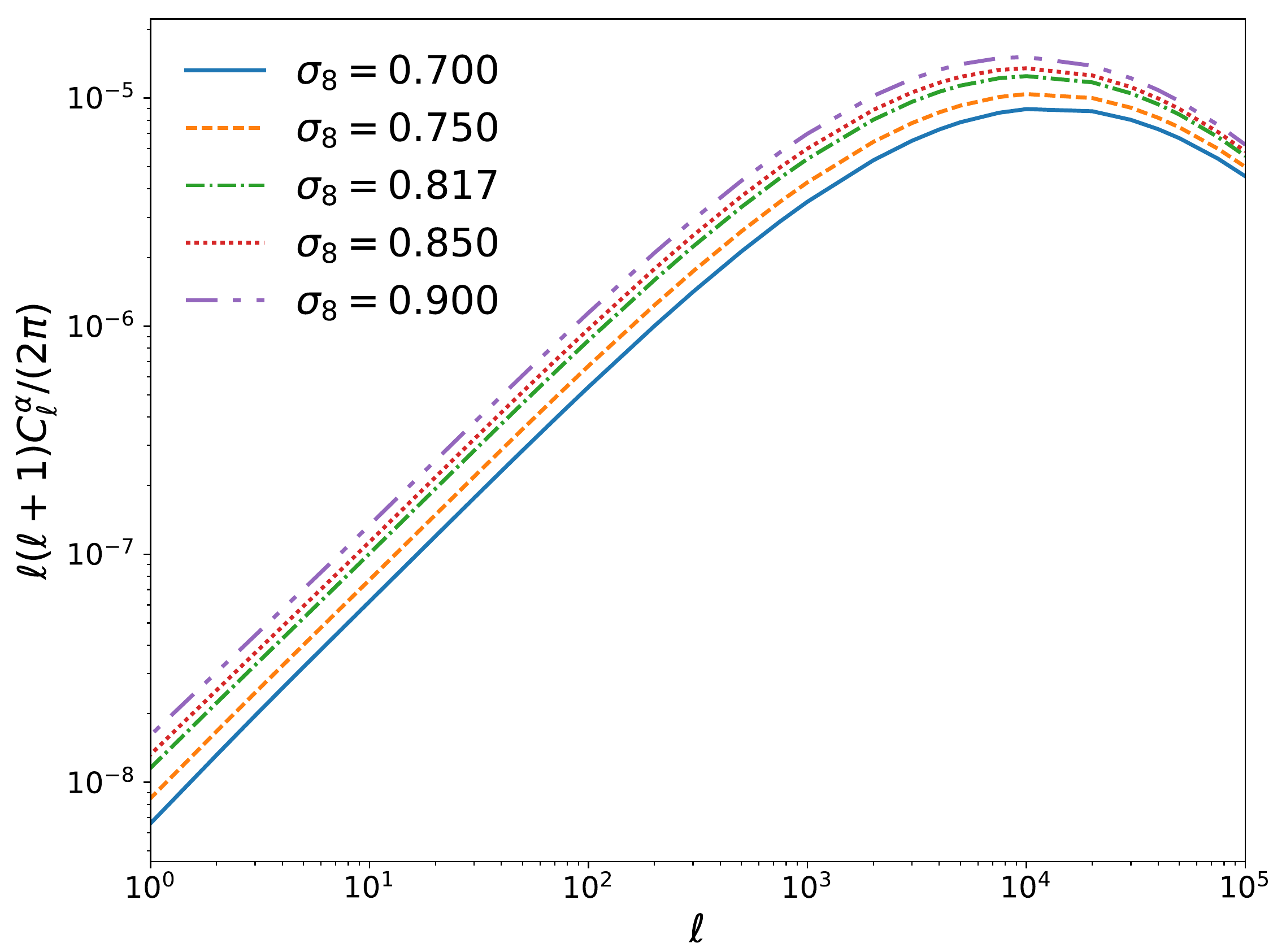}
\includegraphics[scale=0.4]{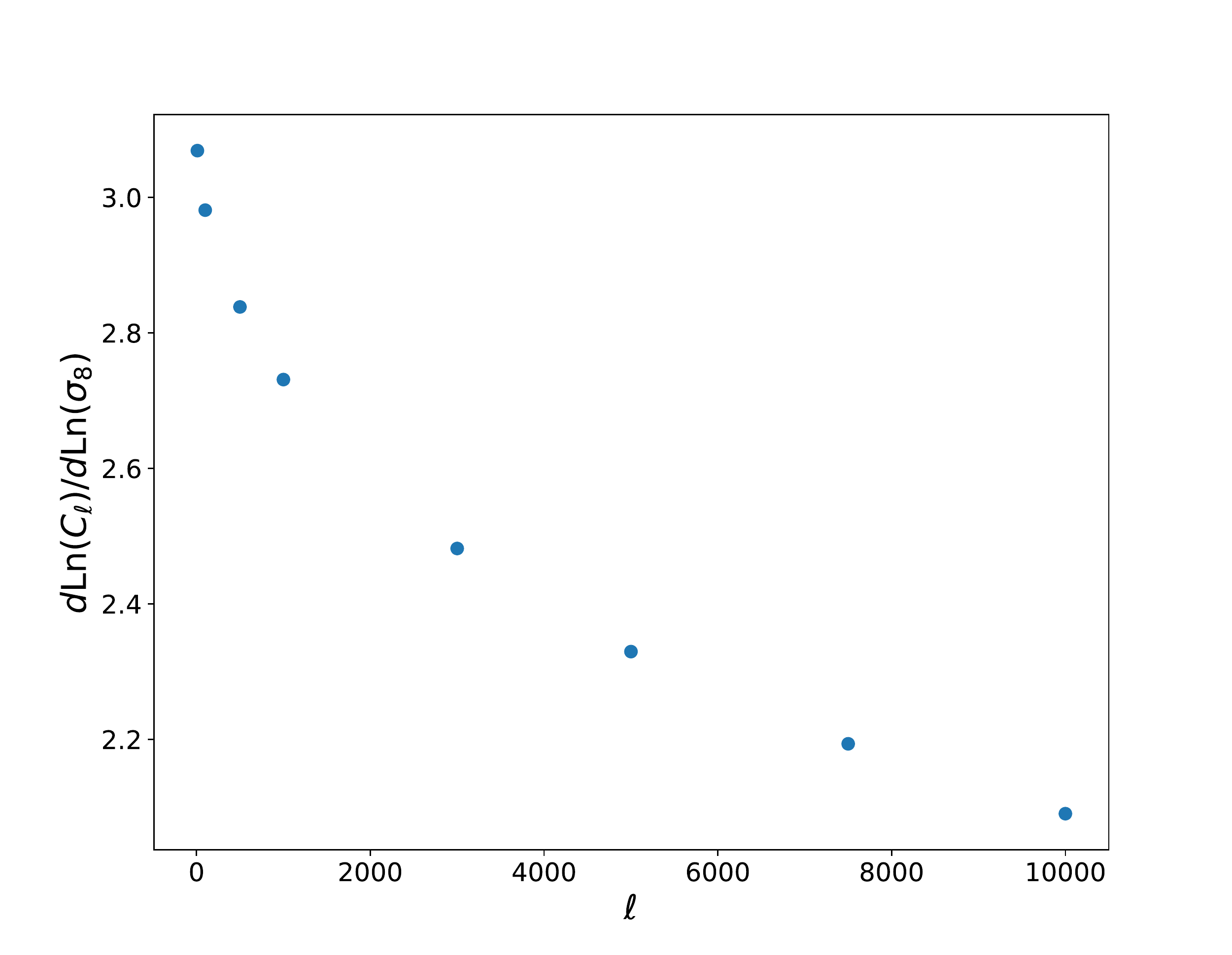}\caption{\label{sigma8_dependence} Left: the angular power spectra of the Faraday rotation angle, $C^\alpha_\ell$, for different values of the density fluctuations amplitude $\sigma_8$. Right: $\dd\ln C^\alpha_\ell/\dd\ln\sigma_8$ as a function of $\ell$}
\end{center}
\end{figure*}

Fig.~\ref{sigma8_dependence} shows two different representations of the dependence of the angular power spectrum of the Faraday rotation angle on the amplitude of density fluctuation $\sigma_8$: on the left is plotted the angular power spectrum for different values of $\sigma_8$ and on the right we plot the logarithmic derivative of the angular power spectrum with respect to $\sigma_8$ as a function of $\ell$. The latter gives the scaling of $C^\alpha_\ell$ with $\sigma_8$, i.e. by writing $C^\alpha_\ell\propto \sigma_8^{n(\ell)}$ then $n(\ell)=\dd \ln(C^\alpha_\ell)/\dd\ln(\sigma_8)$. 

The angular power spectrum $C^\alpha_\ell$ is composed of the 1-halo term only. Hence its scaling with $\sigma_8$ is driven by the mass function, $\dd N/\dd M$, and the rotation angle at the core of halos, $\alpha_c$. The latter does not explicitly depends on $\sigma_8$. However, the scaling of $\dd N/\dd M$ with $\sigma_8$ is mass-dependent. Then the mass dependence of $\alpha_c$ probes different mass ranges of the mass function, and as a consequence, different scaling of $\dd N/\dd M$ with the amplitude of matter perturbations. 

We find a dependence as $C^{\alpha}_\ell\propto \sigma_8^{3.1}-\sigma_8^{2.1}$ for $\ell=10$ and $\ell=10^4$, respectively. The power spectrum of the Faraday rotation angle is more sensitive to $\sigma_8$ for low $\ell$ values than for high $\ell$ values because as seen above, the angular power spectrum is sensitive to higher mass at low $\ell$ and in this mass regime the mass function is more sensitive to $\sigma_8$. We noticed that reducing the mass integration range from $M=10^{13}\ensuremath{M_{\odot}}$ to $M=5\times10^{16} \ensuremath{M_{\odot}}$ (where it was $[10^{10} \ensuremath{M_{\odot}},5\times10^{16} \ensuremath{M_{\odot}}]$ before) slightly increases the power in $\sigma_8$. This may be due both to the facts that the Faraday rotation effect is mainly sensitive to galaxy halos with masses in the range $M=10^{13}$ to $M=10^{15} \ensuremath{M_{\odot}}$, see Fig.~\ref{mass_redshift_dependence}, and that our mass function depends on $\sigma_8$ more strongly from $M=10^{14} \ensuremath{M_{\odot}}$. 

We note that the scaling in $\sigma_8$ of the angular power spectrum is different than the one for the tSZ angular power spectrum which scales with $\sigma_8^{8.1}$ \citep[see for example][]{Hurier:2017jgi}. The reason is a different scaling in mass of the rotation angle at the core of halos as compared to the tSZ flux (note that the tSZ angular power spectrum is dominated by the 1-halo contribution). Indeed, $\left|\alpha_c\right|^2$ scales as $M^2$ whereas the square of the tSZ flux at the core scales as $M^{3.5}$. This results in a different weighting of the mass function which is more sensitive to $\sigma_8$ for high mass values, the tSZ effect giving more weight to high masses than the Faraday rotation angle.

The dependence with $\sigma_8$ found here is however different from the one reported in \citet{Tashiro:2007mf}, the difference being mainly due to the presence of a 2-halo term in \citet{Tashiro:2007mf}. The mass range $\left[M=10^{13}\ensuremath{M_{\odot}},5\times10^{16} \ensuremath{M_{\odot}}\right]$ is first considered in \citet{Tashiro:2007mf} for which the angular power spectrum is dominated by its 1-halo contribution.\footnote{We remind that the angular power spectra derived in \citet{Tashiro:2007mf} has a non zero 2-halo contribution.} In this case, the obtained scaling is $\sigma_8^5$. The difference with the scaling found here lies in the reduced mass range which gives more weight on the total effect to higher mass halos. Second the mass range is extended in \citet{Tashiro:2007mf} down to $10^{11}\ensuremath{M_{\odot}}$, leading then to a scaling as $\sigma_8^{5.5}$. In the mass range $[10^{11}\ensuremath{M_{\odot}},10^{13}\ensuremath{M_{\odot}}]$, the 2-halo term present in \citet{Tashiro:2007mf} is not negligible anymore. This 2-halo term then gives much more contribution to low mass halos as compared to ours \citep[see Fig. 7 of][]{Tashiro:2007mf}. However, the scaling of the 2-halo term with $\sigma_8$ is not driven anymore by $$\sim\int\dd M \frac{\dd N}{\dd M}\alpha_c^2,$$ but instead by $$\sim\left(\int\dd M \frac{\dd N}{\dd M}b(M,z)\alpha_c\right)^2P_\mathrm{m}(\ell/r,z),$$ with $P_\mathrm{m}(k,z)$ being the matter power spectrum (proportional to $\sigma_8$). The steeper scaling with $\sigma_8$ found in \citet{Tashiro:2007mf} is thus mainly due to the non-negligible contribution of the 2-halo term in their work.  \\


We now want to study whether the Faraday rotation angle is sensitive to the matter density parameters. Keeping other cosmological parameters fixed, we have two possibilities to vary $\Omega_m$: either by varying the density of cold dark matter, $\Omega_{CDM}$, or that of baryons, $\Omega_b$.

We found that the Faraday rotation effect is almost independent of $\Omega_m$, when $\Omega_b$ is kept fixed while varying $\Omega_{CDM} $: 
$$C_\ell^{\alpha}\propto \Omega_{CDM}^{-0.1}-\Omega_{CDM}^{-0.2}$$
for $\ell=10$ and $\ell=10^4$ respectively. This translates into a similar scaling with $\Omega_m$ for a varying density of dark matter
$$C_\ell^{\alpha}\propto \Omega_m^{-0.1}-\Omega_m^{-0.2}$$
for $\ell=10$ and $\ell=10^4$ respectively.

However, when keeping $\Omega_{CDM}$ fixed and varying $\Omega_b$, the dependence is clearly different:
$$C_\ell^{\alpha}\propto \Omega_b^{2.0}-\Omega_b^{1.9}$$
for $\ell=10$ and $\ell=10^4$ respectively. The resulting scaling with $\Omega_m$ by varying the density of baryons is then 
$$C_\ell^{\alpha}\propto \Omega_m^{13}-\Omega_m^{12}$$
for $\ell=10$ and $\ell=10^4$ respectively.
The dependence with $\Omega_b$ and $\Omega_{CDM}$ is simply understood by the fact the angular power spectrum scales with the fraction of baryons to the square. The effect is almost $\Omega_m$-independent when varying $\Omega_{CDM}$, as compared to the thermal Sunyaev-Zel'dovich effect which scales as $\sim \Omega_m^3$ \citep{Komatsu:1999ev}.
One can thus hope to use the Faraday rotation as a cosmological probe, by combining it with another physical effect having a different degeneracy in the $\Omega_m-\sigma_8$ plane, such as the thermal Sunyaev-Zel'dovich effect. \\

Finally, we also want to study the effect of having a mass dependence of the (central) magnetic field strength. Indeed, the structure of magnetic fields in galaxy halos is poorly known and this fact could lead to degeneracies in the dependence with cosmological parameters and astrophysical ones. We chose the central magnetic field strength to scale as $B_0=B_p (M/M_p)^{\gamma}$, with $M_p=5\times10^{14}\ensuremath{M_{\odot}}$, $B_P=3$ $\mu$G, and $\gamma$ varying from 0 to 1. 

\begin{figure*}
\begin{center}
\includegraphics[scale=0.4]{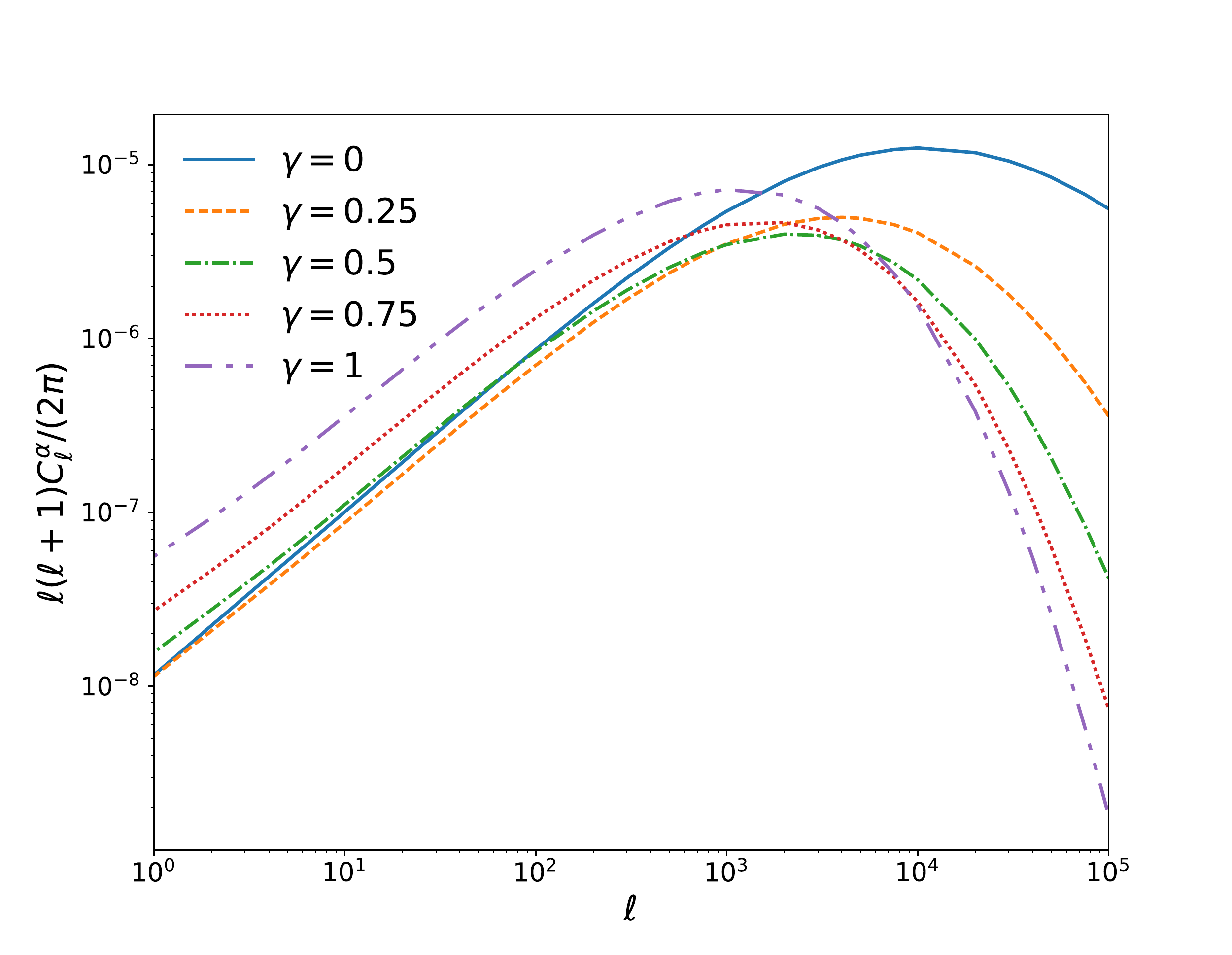}
\includegraphics[scale=0.4]{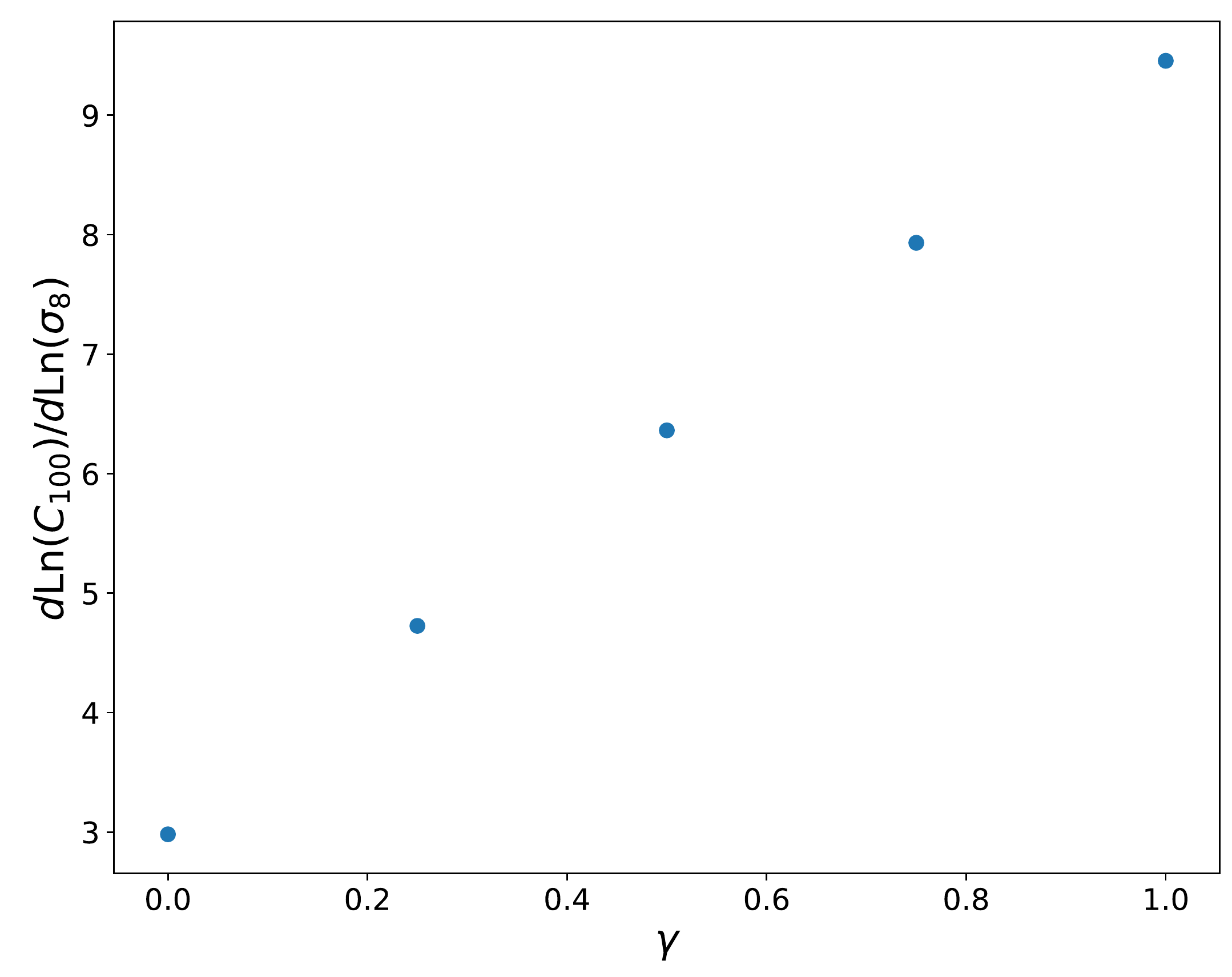}
\caption{\label{mass_dependent_magnetic_field} Left: The angular power spectra of the Faraday rotation angle, $C^\alpha_\ell$, when adding a mass dependence for the magnetic field strength at the centre scaling as $B=B_p(M/M_p)^{\gamma}$, with $M_p=5\times10^{14}\ensuremath{M_{\odot}}$, $B_p=3$ $\mu$G, and for different values of $\gamma$. Right: $\dd\ln C^\alpha_{100}/\dd\ln\sigma_8$ as a function of $\gamma$. We chose to plot this effect for $\ell=100$ as $C^\alpha_\ell$ depends more strongly on $\sigma_8$ for low $\ell$ values.}
\end{center}
\end{figure*}

The left part of Fig.~\ref{mass_dependent_magnetic_field} shows $C^\alpha_\ell$ for five different values of $\gamma$. When increasing $\gamma$, the power spectrum is increased and the peak is shifted to lower $\ell$ values. Increasing the value of $\gamma$ indeed leads to a higher contribution of massive halos, which appears larger once projected on the sky, hence a peak at smaller multipoles. This shift to lower $\ell$ values and the difference in amplitude of the Faraday rotation angle could give insight on the scaling of the magnetic field strength with mass. 

Fig.~\ref{mass_dependent_magnetic_field} right shows how this mass dependence affects the dependence on $\sigma_8$ of the Faraday rotation effect by plotting $\dd\ln C^\alpha_{100}/\dd\ln\sigma_8$ with respect to $\gamma$. For $\ell=100$, when $\gamma=1$, we find $C^{\alpha}_\ell\propto \sigma_8^{9.5}$ and we recover $C^{\alpha}_\ell\propto \sigma_8^{3.0}$ for $\gamma=0$. In between, one has $C^{\alpha}_\ell\propto \sigma_8^{4.7}-\sigma_8^{6.4}-\sigma_8^{7.9}$ for $\gamma=0.25,\ 0.5,\ 0.75$ respectively. We stated few lines above that our different scaling with $\sigma_8$ of the angular power spectrum as compared to the thermal SZ effect came from a different scaling in mass. Indeed, the angular power spectrum of our effect scales as $M^2$, where it scales as $M^{3.5}$ for the tSZ effect, hence we recover the same scaling in $\sigma_8$ for $\gamma=0.75$. From this we also see that if we could model the magnetic field with a power-law mass dependence, the more it would depend on mass, the more the effect would be sensitive to $\sigma_8$, allowing for a better determination of this cosmological parameter. Hence there is a correlation between the uncertainty on $\sigma_8$ and the mass dependence of the magnetic field strength. The Faraday rotation angle still almost does not depend on $\Omega_m$ (when varying $\Omega_{CDM}$ only). Indeed, one has: $C^{\alpha}_\ell\propto \Omega_m^{-0.1}-\Omega_m^{-0.0}-\Omega_m^{-0.0}-\Omega_m^{-0.1}-\Omega_m^{-0.1}$ for $\gamma=0,\ 0.25,\ 0.5,\ 0.75$ and 1 respectively.

\subsection{Power spectra of the Faraday Conversion rate}

\begin{figure}
\begin{center}
\includegraphics[scale=0.35]{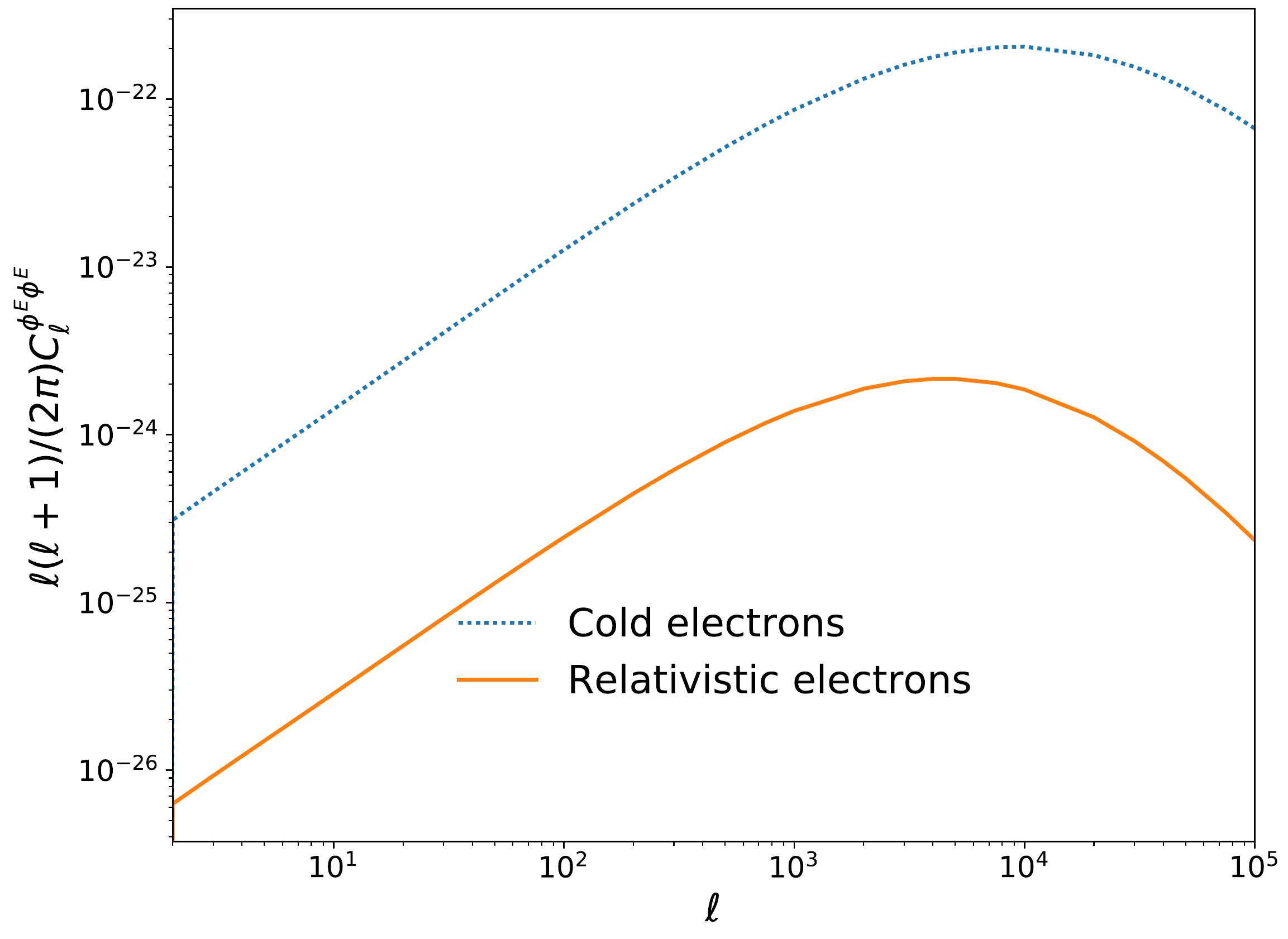}
\caption{\label{FC_comparison} The angular power spectra of the Faraday conversion rate $C^{\phi^{E}\phi^{E}}_\ell$ for thermal electrons (dotted-blue) and for relativistic electrons (solid-orange). For relativistic electrons, one set $n_e^{(r)}=10$ m$^{-3}$ and $\Gamma_\mathrm{min}=300$.}
\end{center}
\end{figure}

First, we found no significative difference between the two angular power spectra of the Faraday conversion rate, $C^{\phi^{E}\phi^{E}}_\ell$ and $C^{\phi^{B}\phi^{B}}_\ell$. Hence, for simplicity we will now show results for the $C^{\phi^{E}\phi^{E}}_\ell$ power spectrum only. \\

On Fig.~\ref{FC_comparison}, we compare the angular power spectra of the Faraday conversion rate for two populations of free electrons, either thermal or relativistic. In the case of the relativistic electrons, the central density in the halo is taken to be constant, contrary to the cold case which follows Eq.~(\ref{concentration}). We took the value $n_e^{(c)}=10$ m$^{-3}$ which is the highest value we could find in the literature \citep{Colafrancesco:2002zq}, noting that the properties of relativistic electrons in halos are not well known. We also took the spectral index of the energy distribution of these relativistic electrons to be $\beta_E=2.5$ and a minimum Lorentz factor of $\Gamma_{min}=300$, which are values typically found in the literature \citep{Feretti:2003gn}. For such values describing relativistic electrons, we find that $C^{\phi^{E}\phi^{E}}_\ell$ is two orders of magnitude higher in the thermal electron case compared to the relativistic case, despite pushing the relativistic electron density to the maximum value allowed by observations.  

We note however that some studies suggest that the minimum Lorentz factor could be as high as $10^4$. Since the angular power spectrum scales in amplitude as $\Gamma^2_\mathrm{min}$, $C^{\phi^E\phi^E}_\ell$ would be two orders of magnitude higher than the one displayed in Fig. \ref{FC_comparison}, hence reaching a similar amplitude as the contribution of thermal electrons to Faraday conversion. One thus expect the contribution of relativistic electrons to be at most at the same amplitude as the one from thermal electrons.

\subsubsection{Thermal electrons}
The dependence of the angular power spectra of the Faraday conversion rate from thermal electrons on the density fluctuation amplitude $\sigma_8$ is similar to the one of the Faraday rotation angle: $C^{\phi^{E}\phi^{E}}_\ell\propto \sigma_8^{3.1}-\sigma_8^{1.9}$ for $\ell=10$ and $\ell=10^4$ respectively, the difference between low $\ell$ and high $\ell$ values having already been explained. The small differences with Faraday rotation come from the fact that the Faraday conversion rate angular power spectra scale as $1/(1+z)^6$, where it scales as $1/(1+z)^4$ for the Faraday rotation angle. 

As for the Faraday rotation effect, there is almost no variation of the Faraday conversion rate with $\Omega_m$, when varying $\Omega_{CDM}$ and $\Omega_b$ kept fixed, the dependence going as $$C_\ell^{\phi^{E}\phi^{E}}\propto \Omega_m^{-0.1}-\Omega_m^{-0.2}\quad \mathrm{or}\quad C_\ell^{\phi^{E}\phi^{E}}\propto \Omega_{CDM}^{-0.1}-\Omega_{CDM}^{0.2}$$ for $\ell=10$ and $\ell=10^4$, respectively.

When varying $\Omega_m$ via $\Omega_b$ instead ($\Omega_{CDM}$ kept fixed), the dependence is not very different from Faraday rotation either: $$C_\ell^{\phi^{E}\phi^{E}}\propto \Omega_m^{13}-\Omega_m^{12}\quad \mathrm{or}\quad C_\ell^{\phi^{E}\phi^{E}}\propto \Omega_b^{2.0}-\Omega_b^{1.9}$$ for $\ell=10$ and $\ell=10^4$, respectively.

Although these scalings are not very different from Faraday rotation, the same remark on the $\sigma_8$ scaling differences applies here, which is that the two effects scale differently with redshift. \\

\begin{figure*}
\begin{center}
\includegraphics[scale=0.35]{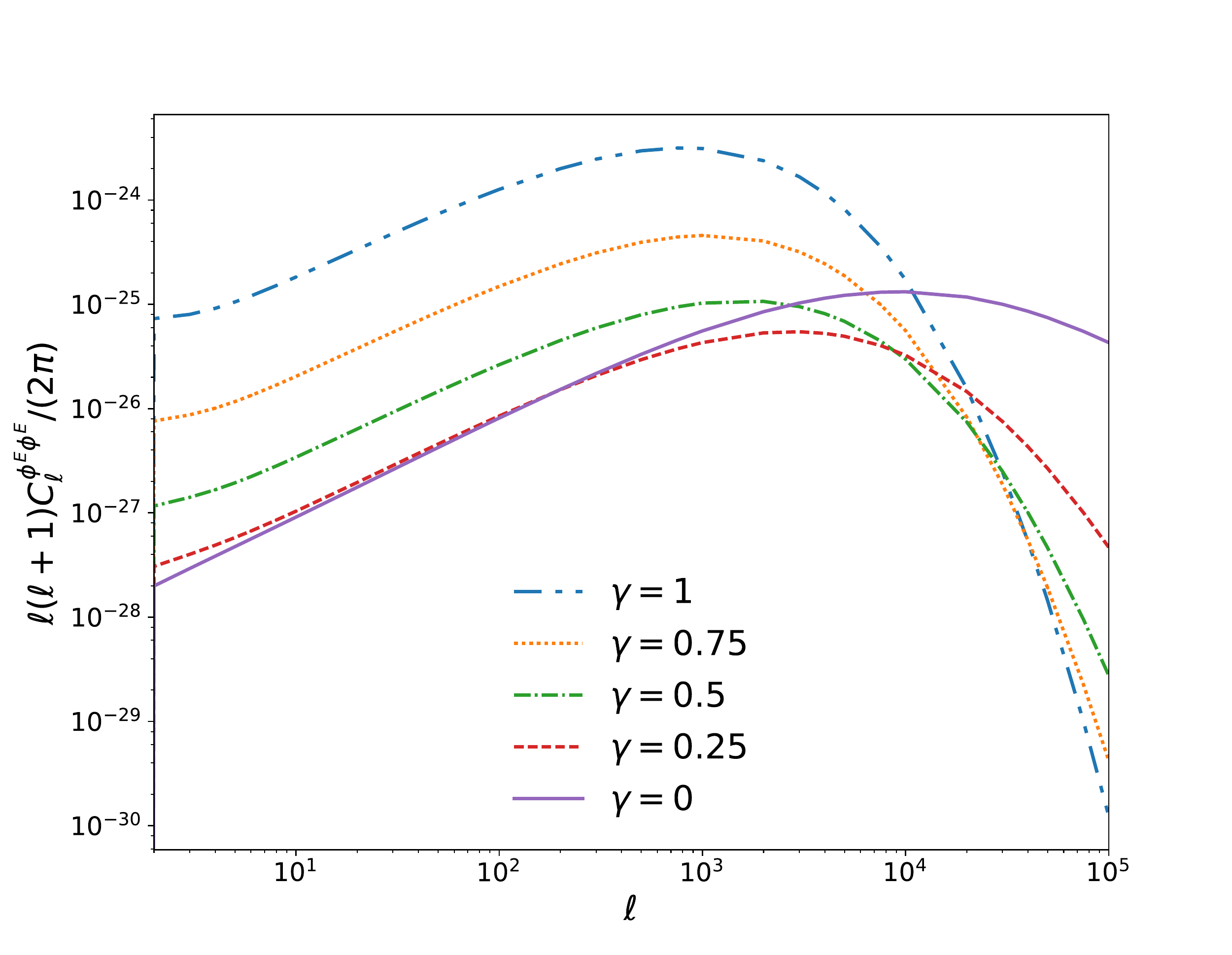}
\includegraphics[scale=0.35]{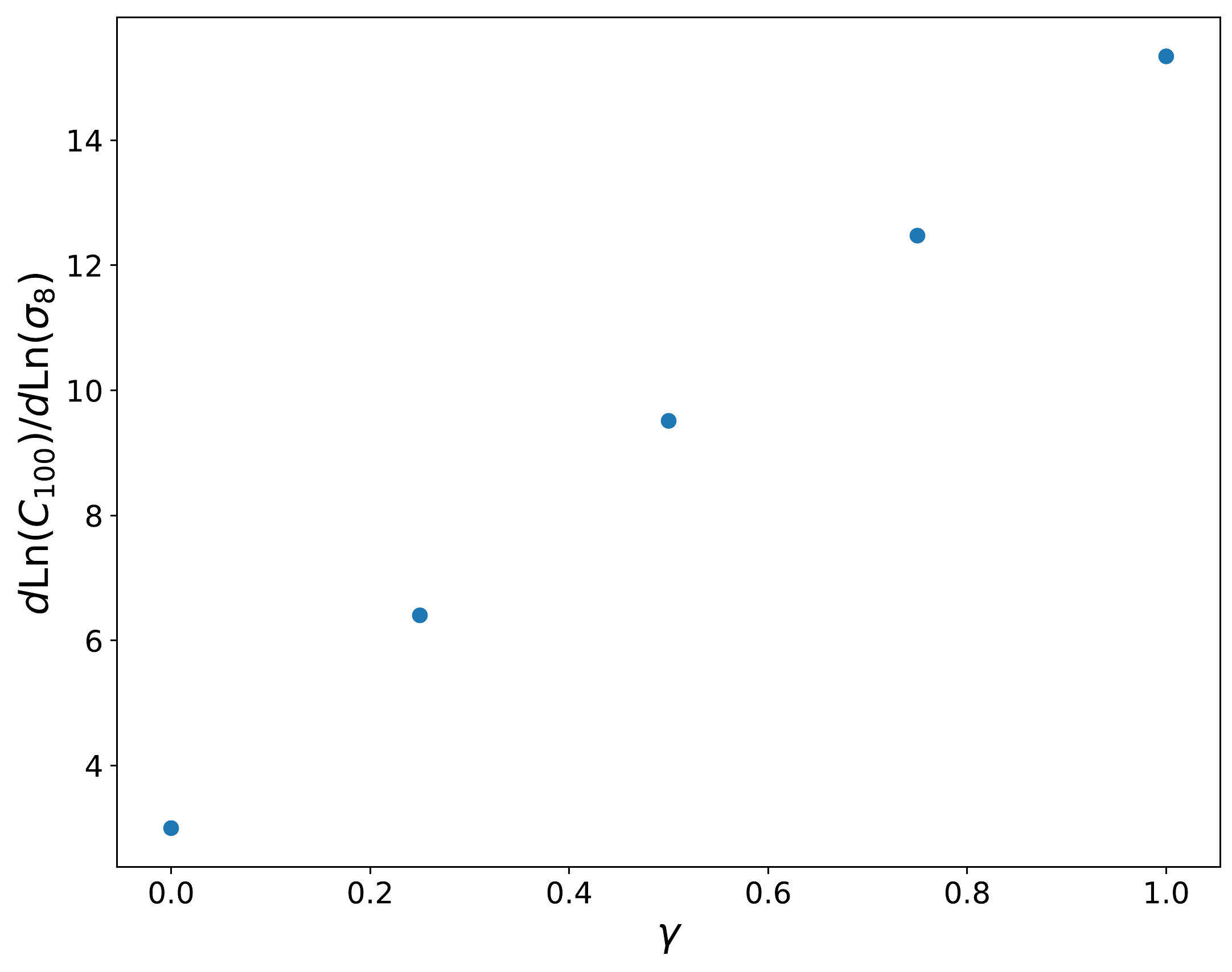}
\caption{\label{mass_dependent_magnetic_field_FC} Left: the angular power spectra of the Faraday conversion rate, $C^{\phi^{E/B}\phi^{E/B}}_\ell$ when adding a mass dependence for the magnetic field strength at the centre scaling as $\sim (M/M_p)^{\gamma}$, with $M_p=5\times10^{13}\ensuremath{M_{\odot}}$ for different values of $\gamma$. Right: $\dd\ln C^\alpha_{100}/\dd\ln\sigma_8$ as a function of $\gamma$. We chose to plot this effect for $\ell=100$ as $C^{\phi^{E/B}\phi^{E/B}}_\ell$ depends more strongly on $\sigma_8$ for low $\ell$ values.}
\end{center}
\end{figure*}

We also investigate the degeneracy between a scaling in mass of the magnetic field at the centre of the halo and the $\sigma_8$ scaling, as what we have done for the Faraday rotation angle, see Fig.~\ref{mass_dependent_magnetic_field_FC}. The dependence in $\sigma_8$ is $C^{\phi^{E}\phi^{E}}_\ell\propto \sigma_8^{15}$ when $\gamma=1$, $C^{\phi^{E}\phi^{E}}_\ell\propto \sigma_8^{12}$ when $\gamma=0.75$, $C^{\phi^{E}\phi^{E}}_\ell\propto \sigma_8^{9.5}$ when $\gamma=0.5$, $C^{\phi^{E}\phi^{E}}_\ell\propto \sigma_8^{6.4}$ when $\gamma=0.25$. When $\gamma=0.5$ the angular power spectra scales with the mass to the four. This results in a scaling with $\sigma_8$ to the power 9.5. For a comparison, it was also 9.5 for the Faraday rotation angle when it scaled with the mass to the four, corresponding for this case to $\gamma=1$, so that our analysis is consistent.

\subsubsection{Relativistic electrons}
Let us mention the case of Faraday conversion with relativistic electrons because the dependence of the angular power spectra with cosmological parameters is a bit different. 

Indeed, when varying $\Omega_{CDM}$ and $\Omega_b$ kept fixed, the dependence goes like: $$C_\ell^{\phi^{E}\phi^{E}}\propto \Omega_m^{1.0}-\Omega_m^{0.7}\quad \mathrm{or}\quad C_\ell^{\phi^{E}\phi^{E}}\propto \Omega_{CDM}^{0.9}-\Omega_{CDM}^{0.6}$$ for $\ell=10$ and $\ell=10^4$, respectively.

When varying $\Omega_{b}$ and $\Omega_{CDM}$ kept fixed, the dependence goes like:
$$C_\ell^{\phi^{E}\phi^{E}}\propto \Omega_m^{0.9284}-\Omega_m^{0.2}\quad \mathrm{or}\quad C_\ell^{\phi^{E}\phi^{E}}\propto \Omega_b^{0.1}-\Omega_b^{0.0}$$ for $\ell=10$ and $\ell=10^4$, respectively.

This difference in dependence compared to the thermal electron case, is explained by the constant value for the density of relativistic free electrons at the centre of the halo, whereas the density of cold free electrons at the centre scales with the fraction of baryons $f_b$ as well as the critical density $\rho_c$ and the spherical overdensity $\Delta$ of the virialized halos. \\

\section{Conclusion}
\label{sec:conclu}
We revisited the derivation of the angular power spectrum of the Faraday rotation angle using the halo model and extended it to the case of Faraday conversion, with an emphasis on the assumptions made for the statistics and orientations of magnetic fields inside halos. Indeed, we first assumed the magnetic field of a halo to have a spherically symmetric profile but the same orientation over the halo scale. Second, the orientations are supposed to be uniformly distributed in the Universe, to be consistent with the cosmological principle. Third, the orientations of magnetic fields in different halos are here independent from each other, with the underlying idea that magnetism is produced within halo in a local physical process. We also made the hypothesis that the distribution of the orientations of the magnetic fields inside halos are independent from the abundance in mass and the spatial distribution of halos. All of these hypotheses simplified the derivation of the angular power spectra: in particular, only the 1-halo term remains because of the independence of the orientations from one halo to another. \\

\begin{table*}
\begin{center}
\begin{tabular}{l|cc}  \hline\hline
	& $\sigma_8$ & $\Omega_m$ \\ \hline
	Faraday rotation power spectrum at $\ell\simeq10^4$ & 2.1 & -0.1 \\
	Faraday conversion power spectrum at $\ell\simeq10^4$ (thermal electrons) & 2.1 & -0.1 \\
	Faraday conversion power spectrum at $\ell\simeq10^4$ (relativistic electrons) & 2.1 & 1 \\ \hline
	Halo number counts from thermal SZ& 9 & 3 \\
	thermal SZ power spectrum at $\ell\simeq 3000$ & 8.1 & 3.2 \\ 
	CMB lensing power spectrum at $\ell\simeq 30$ & 2 & 0.5 \\  \hline\hline
\end{tabular}
\caption{Scaling of different large-scale-structure probes with $\sigma_8$ and $\Omega_m$. The scaling reported here is to be understood as $P\propto\sigma^n_8\,\Omega^p_m$ with $P$ any of the considered probe. They are given at the peaking multipole of $\ell(\ell+1)C_\ell$ for the Faraday rotation angle, the Faraday conversion rate, and the tSZ flux, and the peaking multipole of $\ell^2(\ell+1)^2C^{\phi\phi}_\ell$ for the lensing potential.}
\label{tab:scaling}
\end{center}
\end{table*}

We then explored the dependence of the angular power spectra with astrophysical and cosmological parameters. In tab. \ref{tab:scaling}, we report the scaling of the angular power spectra of the Faraday rotation angle and the Faraday conversion rate with the parameters $\sigma_8$ and $\Omega_m$ (assuming here that $\Omega_m$  would vary by a change of the cold dark matter density). We also reported in this table the scaling for three other probes of the large-scale structures, namely halo number counts as observed through the thermal Sunyaev-Zel'dovich effect (tSZ), the tSZ angular power spectrum, and the angular power spectrum of CMB lensing potential.  

In particular, the angular power spectra of both Faraday rotation and Faraday conversion scale with the amplitude of the density fluctuations as $\sigma_8^3$ while it scales with $\sigma_8^8$ to $\sigma^9_8$ for the other probes. However, this scaling with $\sigma_8$ is degenerated with a mass dependent magnetic field.\footnote{The impact of the scaling of $B$ with the mass can be viewed as the equivalent of the mass bias in the analysis of the tSZ number counts and angular power spectrum.} Still different from the tSZ and lensing probes is the scaling with the matter density parameter $\Omega_m$: while for the SZ one has a scaling with $\Omega_m^3$, here it is almost independent of this parameter. Thus the two effects could be combined to lift the degeneracy in the $\sigma_8-\Omega_m$, assuming nonetheless the magnetic field mass dependence model to be known. Conversely, a joint analysis could be used so as to infer the scaling of the magnetic fields with the masses of halos.
\\

Although other physical effects happen in these magnetized plasmas being halos of galaxies, the dominant contributions are from Faraday rotation and Faraday conversion with thermal electrons as stated in Sect. \ref{sec:halos}. Indeed, an estimation of the angular power spectra of secondary anisotropies suggests that at 1 GHz, with a magnetic field of 10 $\mu$G and a density of relativistic electrons of $n_{rel}=10$ m$^{-3}$ for the absorption coefficients, there are eighteen orders of magnitude between the secondary linearly polarized anisotropies produced thanks to Faraday rotation and those produced by absorption of intensity. Similarly, there are nine orders of magnitude between the secondary circularly polarized anisotropies produced thanks to Faraday conversion and those produced by absorption of intensity. These differences in orders of magnitude do not change significantly when changing the frequency to 30 GHz and the magnetic field to 3 $\mu$G. Thus, we can safely conclude the secondary anisotropies induced by absorption would be negligible compared to the Faraday rotation and Faraday conversion induced anisotropies. \\

\paragraph{Note added:} During the completion of this work, another study of Faraday conversion in the cosmological context has been proposed in \citet{Ejlli:2018ucq}. It however focuses on magnetic fields averaged over very large scales.

\begin{acknowledgements}
The authors would like to thank G. Fabbian, S. Ilic, and M. Douspis for helpful discussions.
Part of the research described in this paper was carried out at the Jet Propulsion Laboratory, California Institute of Technology, under a contract with the National Aeronautics and Space Administration (NASA).
\end{acknowledgements}

\bibliographystyle{aa}
\bibliography{biblio}

\onecolumn
\appendix

\section{Formal derivation of the angular power spectra for any effect considered}
\label{app:wigner}
\subsection{Radiative transfer coefficients}
Looking at the expressions of the different radiative transfer coefficients show that they can assume two possible forms. They are first of scalar type as the Faraday rotation angle, and they read as an integral over the line-of-sight and over halos distribution as follows
\begin{equation}
	\alpha(\vec{n})=\ds\int_0^{r_{\mathrm{CMB}}} a(r)\dd r \iint\dd M_i \dd^3\vec{x}_i~n_h(\vec{x}_i)~ \left[\vec{b}(\vec{x}_i)\cdot\vec{n}\right]A\left(M_i,\left|\vec{x}-\vec{x}_i\right|\right).
\end{equation}
In the above, the dependance over orientations is encoded in $\left[\vec{b}(\vec{x}_i)\cdot\vec{n}\right]$ which is a scalar function. The function $A$ is the profile of the effect which reads for Faraday rotation by thermal electrons
\begin{equation}
	A\left(M_i,\left|\vec{x}-\vec{x}_i\right|\right)=\left(\frac{e^3}{8\pi^2\,m_e^2\,c\,\varepsilon_0\nu^2(r)}\right)n_e\left(\left|\vec{x}-\vec{x}_i\right|\right)B\left(\left|\vec{x}-\vec{x}_i\right|\right).
\end{equation}
Since projection effects can only reduce the impact of the effect, the quantity $A$ can be interpreted as the maximum of the effect a given halo can generate. Finally, $n_h$ is the halo distribution. The above scalar-type of coefficients are the Faraday rotation angle and the conversion from intensity to circular polarization, $\phi^{I\to V}$.

The second type of coefficients are the ones which are proportional to $B^2_\perp e^{\pm 2i\gamma}$, with $B_\perp$ the amplitude of the {\it projected} magnetic field on the plane orthogonal to the line-of-sight, and $\gamma$ the angle between the projected magnetic field with the basis vector $\vec{e}_\theta$ in that plane. This is typical of the coefficients for Faraday conversion, $\phi^{P\to V}$, or conversion from intensity to linear polarization, $\phi^{I\to P}$. In terms of the amplitude of the magnetic field, $B$, and its orientation $\vec{b}$, the phase reads $B^2_\perp e^{\pm2i\gamma}=B~\left[\vec{b}\cdot\left(\vec{e}_\theta\pm i\vec{e}_\varphi\right)\right]^2$. One thus have spin-$(\pm2)$ coefficients reading as an integral over the line-of-sight as follows
\begin{equation}
	i\phi^{P\to V}(\vec{n})e^{\pm2i\gamma(\vec{n})}=\ds\int_0^{r_{\mathrm{CMB}}} a(r)\dd r \iint\dd M_i \dd^3\vec{x}_i~n_h(\vec{x}_i)~ \left[\vec{b}\cdot\left(\vec{e}_\theta\pm i\vec{e}_\varphi\right)\right]^2P\left(M_i,\left|\vec{x}-\vec{x}_i\right|\right).
\end{equation}
The spin structure of the above is entirely encoded in projection coefficients $\left[\vec{b}\cdot\left(\vec{e}_\theta\pm i\vec{e}_\varphi\right)\right]^2$. The other terms are identical to the ones in scalar coefficients, except that the profile of the effect, $P\left(r,\left|\vec{x}-\vec{x}_i\right|\right)$, admits a different explicit expression, e.g. for Faraday conversion by thermal electrons
\begin{equation}
	P\left(M_i,\left|\vec{x}-\vec{x}_i\right|\right)=\left(\frac{e^4}{16\pi^2m^2_ec^3\nu^3(r)}\right)n_e\left(\left|\vec{x}-\vec{x}_i\right|\right)B^2\left(\left|\vec{x}-\vec{x}_i\right|\right).
\end{equation}
A part from the explicit expression of $P$, the formal expression of the coefficient remains the same for relativistic electrons or for conversion from intensity to linear polarization.

A formal expression for all the above radiative transfer coefficients can be abstracted from the above. On denoting $\phi_s$ any such coefficients, with $s=0$ for scalar ones and $s=\pm2$ for the spin ones, it is given by
\begin{equation}
	\phi_s(\vec{n})=\ds\int_0^{r_{\mathrm{CMB}}} a(r)\dd r \iint\dd M_i \dd^3\vec{x}_i ~n_h(\vec{x}_i)~f_s(\vec{b}_i,\vec{n})\Phi\left(M_i,\left|\vec{x}-\vec{x}_i\right|\right),
\end{equation}
with $\Phi\left(r,\left|\vec{x}-\vec{x}_i\right|\right)$ the profile amounting the maximum amount of the effect, which is a scalar function, and $f_s(\vec{b}_i,\vec{n})$ the function encoding the impact of projecting the magnetic field (the subscript $i$ is to remind that the orientation is a priori a function of the halos positions, $\vec{x}_i$). This is this last function which contains the spin structure of the considered coefficients. 

\subsection{Angular power spectrum}
To compute the angular power spectrum, the usual approach consists in first computing the multipolar coefficients of $\phi_s(\vec{n})$ thanks to $\phi_{s,\ell,m}=\int\dd \vec{n}\phi_s(\vec{n}){}_sY^\star_{\ell m}(\vec{n})$, and then to consider the 2-point correlation between these multipolar coefficients, $\left<\phi^{(1)}_{s,\ell m}\phi^{(2)~\star}_{s',\ell'm'}\right>$ (superscripts $1,~2$ labels two possibly different radiative transfer coefficients). The fields $\phi_s$ being statistically homogeneous and isotropic, the 2-point correlation of multipolar coefficients is entirely described by an angular power spectrum, i.e. $\left<\phi^{(1)}_{s,\ell m}\phi^{(2)~\star}_{s',\ell'm'}\right>=C^{(1,2)}_\ell~\delta_{\ell,\ell'}\delta_{m,m'}$. 

Here we adopt a slightly different path (totally equivalent though) by first considering the 2-point correlation function on the sphere, denoted $\xi_{1,2}(\vec{n}_1,\vec{n}_2)=\left<\phi_s^{(1)}(\vec{n}_1)\phi_{s'}^{(2)}(\vec{n}_2)\right>$, which is further simplified thanks to our assumption about the statistics of the magnetic fields orientations. The 2-point correlation of the multipolar coefficients is secondly derived from the 2-point correlation function via $\left<\phi_{s,\ell m}\phi^\star_{s',\ell'm'}\right>=\int\dd\vec{n}_1\int\dd\vec{n}_2~\xi_{1,2}(\vec{n}_1,\vec{n}_2){}_sY^\star_{\ell m}(\vec{n}_1){}_{s'}Y_{\ell m}(\vec{n}_2)$. \\

 Assuming that orientations of the magnetic fields is not correlated to the spatial distribution of halos leads to
\begin{eqnarray}
	\xi_{1,2}(\vec{n}_1,\vec{n}_2)&=&\ds\int_0^{r_{\mathrm{CMB}}} \left[a(r_1)\dd r_1\right]\left[a(r_2)\dd r_2\right]\iint\left[\dd M_i\dd^3\vec{x}_i\right]\left[\dd M_j\dd^3\vec{x}_j\right]\Phi^{(1)}\left(M_i,\left|\vec{x}_1-\vec{x}_i\right|\right)\Phi^{(2)}\left(M_j,\left|\vec{x}_2-\vec{x}_j\right|\right) \nonumber \\
	&&\times\left<n_h(\vec{x}_i)n_h(\vec{x}_j)\right>\left<f_s(\vec{b}_i,\vec{n}_1)f_{s'}(\vec{b}_j,\vec{n}_2)\right>.
\end{eqnarray}
Since the orientation of the magnetic fields of two different halos is uncorrelated, this gives $\left<f_s(\vec{b}_i,\vec{n}_1)f_{s'}(\vec{b}_j,\vec{n}_2)\right>\propto\delta_{i,j}$, and only the 1-halo term contributes to the 2-point cross-correlation function. In addition, these orientations are statistically homogeneous and isotropic, meaning that $\left<f_s(\vec{b}_i,\vec{n}_1)f_{s'}(\vec{b}_j,\vec{n}_2)\right>$ is a function of $\left|\vec{x}_i-\vec{x}_j\right|$ only. Because there is only the 1-halo term, this relative distance is zero and $\left<f_s(\vec{b}_i,\vec{n}_1)f_{s'}(\vec{b}_j,\vec{n}_2)\right>$ is a function of $\vec{n}_1$ and $\vec{n}_2$, i.e. $\left<f_s(\vec{b}_i,\vec{n}_1)f_{s'}(\vec{b}_j,\vec{n}_2)\right>=\xi^O_{s,s'}(\vec{n}_1,\vec{n}_2)\,\delta_{i,j}$.  Hence the 2-point correlation function boils down to
\begin{equation}
	\xi_{1,2}(\vec{n}_1,\vec{n}_2)=\xi^O_{s,s'}(\vec{n}_1,\vec{n}_2)\,\times\,\ds\int_0^{r_{\mathrm{CMB}}} \left[a(r_1)\dd r_1\right]\left[a(r_2)\dd r_2\right]\iint\dd M_i\dd^3\vec{x}_i\frac{\dd N}{\dd M}\Phi^{(1)}\left(M_i,\left|\vec{x}_1-\vec{x}_i\right|\right)\Phi^{(2)}\left(M_i,\left|\vec{x}_2-\vec{x}_i\right|\right),
\end{equation}
where the mass function arises from the 1-halo average of the abundance $\left<n^2_h(\vec{x}_i)\right>=\dd N/\dd M$. The function $\xi^O_{s,s'}(\vec{n}_1,\vec{n}_2)$ is interpreted as the correlation function of orientations, while the remaining term is the 1-halo contribution of the 2-point correlation function of the  amplitude of the radiative transfer coefficient. Let us denote this second correlation function $\xi_{1,2}^{\Phi}$. \\

The full correlation function is thus a product of two correlation functions, one for the orientation and one for the amplitude of the coefficient, i.e.
\begin{equation}
	\xi_{1,2}(\vec{n}_1,\vec{n}_2)=\xi^O_{s,s'}(\vec{n}_1,\vec{n}_2)\times \xi_{1,2}^{\Phi}(\vec{n}_1,\vec{n}_2). \label{eq:fullcorr}
\end{equation}
The correlation function of the amplitude, $\xi_{1,2}^{\Phi}$, is formally identical to the 1-halo term of the correlation function of e.g. the thermal Sunyaev-Zel'dovich effect, which is well-known to be described by an angular power spectrum, i.e.
\begin{equation}
	\xi_{1,2}^{\Phi}(\vec{n}_1,\vec{n}_2)=\ds\sum_{L,M}D^\Phi_L~Y_{LM}(\vec{n}_1)Y^\star_{LM}(\vec{n}_2), \label{eq:ampcorr}
\end{equation}
with $D^\Phi_L$ the angular power spectrum. Similarly, the correlation function of orientations is described by an angular power spectrum, $D^O_{L'}$, since this is a statisically homogeneous and isotropic field, i.e.
\begin{equation}
	\xi^O_{s,s'}(\vec{n}_1,\vec{n}_2)=\ds\sum_{L'M'}D^O_{L'}~{}_{s}Y_{L'M'}(\vec{n}_1){}_{s'}Y^\star_{L'M'}(\vec{n}_2). \label{eq:orientcorr}
\end{equation}
We note that in the above, spin-weighted spherical harmonics are used to take into account the nonzero spins of the projected orientations.

Plugging Eqs. (\ref{eq:ampcorr}) \& (\ref{eq:orientcorr}) into Eq. (\ref{eq:fullcorr}), and then taking the spherical harmonic transforms of $\xi_{1,2}$, one shows that $\left<\phi^{(1)}_{s,\ell m}\phi^{(2)~\star}_{s',\ell'm'}\right>$ can be expressed as a function of Gaunt integrals, the latter being defined as 
\begin{eqnarray}
	G^{\ell_1m_1s_1}_{\ell_2m_2s_2;\ell_3m_3s_3}=\displaystyle\int\dd\vec{\hat{n}}\,{_{s_1}Y}_{\ell_1m_1}(\vec{\hat{n}})\,\times\,{_{s_2}Y}_{\ell_2m_2}(\vec{\hat{n}})\,\times\,{_{s_3}Y}_{\ell_3m_3}(\vec{\hat{n}}).
\end{eqnarray}
Gaunt integrals can be casted as products of Wigner-$3j$ symbols. By then using triangular conditions and symmetries of the Wigner symbols \citep{varshalovich1988quantum}, one finds
\begin{align}
	\left<\phi^{(1)}_{s,\ell m}\phi^{(2)~\star}_{s',\ell'm'}\right>=\frac{\sqrt{(2\ell+1)(2\ell'+1)}}{4\pi}\,\displaystyle\sum_{L,L'}(2L+1)(2L'+1)&\,D^\Phi_L~D^O_{L'}\,
	\left(\begin{array}{ccc}
		L' & L & \ell \\
		-s & 0 & s
		\end{array}\right)\left(\begin{array}{ccc}
		L' & L & \ell' \\
		-s' & 0 & s'
		\end{array}\right)\\ \nonumber
		&\displaystyle\sum_{M,M'}\left(\begin{array}{ccc}
		L'& L & \ell \\
		M' & M & -m
		\end{array}\right)\left(\begin{array}{ccc}
		L' & L & \ell' \\
		M' & M & -m'
		\end{array}\right).
\end{align}
The last summation over $M$ and $M'$ of two Wigner-$3j$'s is equal to $(2\ell+1)^{-1}\,\delta_{\ell,\ell'}\,\delta_{m,m'}$. One thus finally obtains that the correlation matrix of the multipolar coefficients is diagonal (as expected for statistically homogeneous and isotropic process), i.e. 
\begin{equation}
	\left<\phi^{(1)}_{s,\ell m}\phi^{(2)~\star}_{s',\ell'm'}\right>=C^{(1,2)}_{\ell}\,\delta_{\ell,\ell'}\,\delta_{m,m'},
\end{equation}
with the angular power spectrum of the Faraday effect given by
\begin{eqnarray}
	C^{(1,2)}_{\ell}=\frac{1}{4\pi}\,\displaystyle\sum_{L,L'}\,(2L+1) (2L'+1)\, \left(\begin{array}{ccc}
		L' & L & \ell \\
		-s & 0 & s
		\end{array}\right)\left(\begin{array}{ccc}
		\ell & L & \ell \\
		-s' & 0 & s'
		\end{array}\right)
        ~D^\Phi_L \ D^O_{L'}.
\end{eqnarray}
Since the 2-point correlation function is the product of two 2-point correlation functions, we consistently find that the angular power spectrum is the convolution of the respective two angular power spectra $D^\Phi_L$ and $D^O_{L'}$. 

\section{Derivation of $D^A_L$}
\label{app:dl}
We describe the derivation of the expression of $D??_\ell$. This is very reminiscent to the calculation of the angular power spectrum of e.g. the thermal Sunyaev-Zel'dovich effect \citep[see for example][]{1988MNRAS.233..637C,1993ApJ...405....1M,Komatsu:1999ev}, here simplified since one only need to derive the 1-halo term. To this end let us define $A(\vec{\hat{n}})$ such that:
\begin{eqnarray}
	A(\vec{\hat{n}})=\frac{e^3}{8\pi^2\,m_e^2\,c\,\varepsilon_0}\ds\int_0^{r_{\mathrm{CMB}}}\frac{a(r)\dd r}{\nu^2(r)}\,\iint \dd M_i \dd^3\vec{x}_i n_h(\vec{x}_i)  X\left(\left|\vec{x}-\vec{x}_i\right|\right).
\end{eqnarray}
Then $D^A_L$ is the angular power spectrum of the above quantity restricted to its 1-halo contribution.

The integral over $\vec{x}_i$ in $A(\vec{n})$ is the convolution of the halo abundance, $n_h$, with the profile of the halo, $X$. This is then written as a product in Fourier space to get
\begin{equation}
	A(\vec{n})= \frac{e^3}{8\pi^2\,m_e^2\,c\,\varepsilon_0}\ds\int_0^{r_{\mathrm{CMB}}}\frac{a(r)\dd r}{\nu^2(r)}\,\iint \dd M_i \,\dd^3\vec{k}\ \widetilde n_{h}(\vec{k},M_i)\,\widetilde{X}(\vec{k})e^{i\vec{k}\cdot\vec{x}},
\end{equation}
with $\tilde{f}(\vec{k})$ meaning the 3D Fourier transform of $f(\vec{x}_i)$. The radial profile being spherically symmetric, it only depends on the norm of the wavevector $k\equiv\left|\vec{k}\right\vert$ and can be expressed using spherical Bessel functions
\begin{equation}
	\widetilde{X}(\vec{k})=\widetilde{X}(k)=\displaystyle\sqrt{\frac{2}{\pi}}\int_0^\infty \dd R~R^2~X(R)\,j_0(kR),
\end{equation}
with $R\equiv\left|\vec{x}-\vec{x}_i\right|$ and $j_0$ the spherical Bessel function at order $\ell=0$. We further make use of the Rayleigh formula to express the $e^{i\vec{k}\cdot\vec{x}}$ using spherical Bessel functions and spherical harmonics. The multipolar coefficients are then obtained through $A_{LM}=\int\dd\vec{n}~A(\vec{n})Y^\star_{LM}(\vec{n})$ leading to
\begin{eqnarray}
	A_{LM}=\frac{e^3}{8\pi^2\,m_e^2\,c\,\varepsilon_0}\ds\int_0^{r_{\mathrm{CMB}}}\frac{a(r)\dd r}{\nu^2(r)}\,\iint \dd M_i \,\dd^3\vec{k}\ \widetilde n_{h}(\vec{k},M_i)\,\widetilde{X}({k})\times(4\pi)\displaystyle\sum_{L,M}(i)^L\,j_L(kr)\,Y^\star_{LM}(\vec{k}/k).
\end{eqnarray}
The 2-point correlation of the above set of multipolar coefficients will involve the auto-correlation of the Fourier transform of the halo abundance. The Poisson part of the 2-point correlation of the halo density field reads $\left<n^2_{h}(\vec{x}_i,M_i)\right>=(\dd N/\dd M_i)\,\delta(M_i-M_j)\,\delta^3(\vec{x}_i-\vec{x}_j)$ with $\dd N/\dd M_i$ the mass function. The corresponding power spectrum is constant (independent of scale) : $\left<\widetilde n_{h}(\vec{k})\,\widetilde n^\star_{h}(\vec{q})\right>=(\dd N/\dd M_i)\,\delta(M_i-M_j)\,\delta^3(\vec{k}-\vec{q})$. Thanks to the scale independance of it, and to the fact that the Fourier-transformed profile of the angle depends on $k$ only, one can perform the integral over $(\vec{k}/k)$ to get
\begin{equation}
	\left<A_{LM}A^\star_{L'M'}\right>=D^A_L~\delta_{L,L'}~\delta_{M,M'},
\end{equation}
with the angular power spectrum 
\begin{eqnarray}
	D^A_L&=&\left(\frac{e^3}{2\pi\,m_e^2\,c\,\varepsilon_0}\right)^2\displaystyle\int_0^{r_{\mathrm{CMB}}}\frac{a(r_1)\,\dd r_1}{\nu^{2}(r_1)}\int_0^{r_{\mathrm{CMB}}}\frac{a(r_2)\,\dd r_2}{\nu^{2}(r_2)}\int \dd M\,\frac{\dd N}{\dd M}\int k^2\,\dd k\,\left|\widetilde{X}(k)\right|^2\,j_L(kr_1)\,j_L(kr_2). \label{eq:fulldl}
\end{eqnarray}
\\

The numerical evaluation of the angular power spectrum $D_L$ as derived above is still prohibitive due to the presence of the highly oscillating Bessel functions. It is however built from expressions of the form
$$
\displaystyle\iint\dd r_1\,\dd r_2\,H_1(r_1)\,H_2(r_2)\int\frac{2k^2\,\dd k}{\pi}\,P(k)\,j_\ell(kr_1)\,j_\ell(kr_2),
$$
which can be simplified using the Limber's approximation \citep{LoVerde:2008re}. Using this approximation, we obtain
\begin{eqnarray}
D^A_L=\left(\frac{e^3}{m_e^2\,c\,\varepsilon_0\,\sqrt{8\pi}}\right)^2\displaystyle\int_0^{r_{\mathrm{CMB}}}\dd r \frac{a^2(r)}{r^2\,\nu^4(r)}\int \dd M\,\frac{\dd N}{\dd M}\,\left\vert\widetilde{X}\left(\frac{L+1/2}{r}\right)\right\vert^2.
\end{eqnarray}
We finalize our expression of the angular power spectrum $D^A_L$ by introducing the projected Fourier transform of the profile. To this end, we first note that $X(R)=X^{(c)}(M,z,B_c)\,U(R/R_c)$ with $B_c$ the mean magnetic field strength at the center of the halo (which can also depend on $M$ and $z$, see \citet{Tashiro:2007mf}), and $U$ a normalized profile which only depends on the ratio of the comoving distance from the center, $R$, to the typical comoving radius of the halo, $R_c$, which is also a function of $z$ and $M$. For a $\beta$-profile, they read $X^{(c)}=n^{(c)}_eB_c$ and $U(R/R_c)=(1+R/R_c)^{-3\beta(1+\mu)/2}$. 

Introducing the variable $x=R/R_c$ and physical radius of the halo, $r^{\mathrm{(phys)}}_c=a(z)R_c$, one finds:
\begin{eqnarray}
	\widetilde{X}\left(\frac{\ell+1/2}{r}\right)&=&\displaystyle\left(\frac{r^2}{a(r)}\right)\,X^{(c)} \times\sqrt{\frac{2}{\pi}}\left(\frac{r^{\mathrm{(phys)}}_c}{\ell_c^2}\right)\int_0^\infty U(x)\,j_0\left((\ell+1/2)x/\ell_c\right)\,x^2\,\dd x, \nonumber
\end{eqnarray}
with $\ell_c=D_\mathrm{ang}(z)/r^{\mathrm{(phys)}}_c$ the typical multipole associated to the typical size of the halo (the latter being also a function of $M$ and $z$ through $r^{\mathrm{(phys)}}_c$), and $D_\mathrm{ang}(z)$ the angular diameter distance. By defining the projected Fourier transform of the profiles
\begin{eqnarray}
	\alpha_\ell(M,z)=\sqrt{\frac{2}{\pi}}\,\left(\frac{r^{\mathrm{(phys)}}_c}{\ell_c^2}\right)\int_0^\infty U(x)\,j_0((\ell+1/2)x/\ell_c)\,x^2\,\dd x,
\end{eqnarray}
the angular power spectrum $D_L$ then writes:
\begin{eqnarray}
	D^A_L&=&\left(\frac{e^3}{m_e^2\,c\,\varepsilon_0\,\sqrt{8\pi}}\right)^2\displaystyle\int\frac{\dd z}{\nu^{4}(z)}\frac{\dd r}{\dd z}\,r^2 \int\dd M\,\frac{\dd N}{\dd M}\,\left|X^{(c)}\right|^2\,\alpha^2_L.\label{eq:Dell}
\end{eqnarray}

\section{Derivation of $D^\parallel_{L}$ for Faraday rotation}
\label{app:ofr}
The angular power spectrum $D^\parallel_L$ for Faraday rotation is obtained through the computation of the correlation $\left<b(\vec{n}_1,\vec{x}_i)b(\vec{n}_2,\vec{x}_j)\right>$. (We remind that $b(\vec{n},\vec{x}_i)=\vec{n}\cdot\vec{b}(\vec{x}_i)$.) We will work using the vector basis $(\vec{{e}}_z,\vec{{e}}_+,\vec{{e}}_-)$ where $\vec{{e}}_\pm=(\vec{{e}}_x\pm i\vec{{e}}_y)/\sqrt{2}$, and $(\vec{{e}}_x,\vec{{e}}_y,\vec{{e}}_z)$ is the standard cartesian basis of $\mathbb{R}^3$. The components of the orientation of the magnetic field $\vec{{b}}$ and the line-of-sight direction $\vec{{n}}$ are given by:
\begin{eqnarray}
	\vec{{b}}=\left(\begin{array}{c}
		\cos(\beta(\vec{x}_i)) \\
		\displaystyle\frac{1}{\sqrt{2}}\sin(\beta(\vec{x}_i))e^{i\alpha(\vec{x}_i)} \\
		\displaystyle\frac{-1}{\sqrt{2}}\sin(\beta(\vec{x}_i))e^{-i\alpha(\vec{x}_i)}
	\end{array}\right), & \mathrm{and,}&\vec{{n}}=2\sqrt{\frac{\pi}{3}}\left(\begin{array}{c}
		Y^{0}_1(\vec{{n}}) \\
		Y^{-1}_1(\vec{{n}}) \\
		Y^{1}_1(\vec{{n}})	
	\end{array}\right).
\end{eqnarray}
We note that in the specific reference frame adopted here, the components of the line-of-sight unit vector are expressed using the spherical harmonics for $\ell=1$. This way of expressing the components of the unit vector of the line-of-sight is appropriate for further reading the angular power spectrum from the 2-point correlation function; see Eq. (\ref{eq:orientcorr}).

For uniformly distributed unit vectors, one obtains the following average:
\begin{equation}
	\left<\vec{b}(\vec{x}_i)\vec{b}(\vec{x}_j)\right>=\left(\begin{array}{ccc}
		1/3 & 0 & 0 \\
		0 & 0 & -1/3 \\
		0 & -1/3 & 0
	\end{array}\right)\,\delta_{i,j}~,
\end{equation}
which is only nonzero for the same halos. This is also constant in space because, as explained in App. \ref{app:wigner}, it results from an homogeneous and isotropic process. The 2-point correlation function finally reads
\begin{equation}
	\left<b(\vec{n}_1,\vec{x}_i)b(\vec{n}_2,\vec{x}_j)\right>=\frac{4\pi}{9}\delta_{i,j}\displaystyle\sum_{m=-1}^{1}Y_{1,m}(\vec{\hat{n}}_1)~Y^\star_{1,m}(\vec{\hat{n}}_2),
\end{equation} 
from which the angular power spectrum is easily obtained to be $D^\parallel_L=\left({4\pi}/{9}\right)\,\times\,\delta_{L,1}$.

\section{Derivation of $D^\perp_{L}$ for Faraday conversion}
\label{app:ofc}
In this appendix one computes the following 2-point correlation functions: $\left<b_{\pm2}(\vec{n}_1,\vec{x}_i)b_{\pm2}(\vec{n}_2,\vec{x}_j)\right>$ and $\left<b_{\pm2}(\vec{n}_1,\vec{x}_i)b_{\mp2}(\vec{n}_2,\vec{x}_j)\right>$ where we remind that $b_{\pm2}\left(\vec{n},\vec{x}_i\right)\equiv[\vec{b}(\vec{x}_i)\cdot(\vec{e}_\theta\pm i\vec{e}_\varphi)]^2$. Working in the basis $(\vec{{e}}_z,\vec{{e}}_+,\vec{{e}}_-)$ as used in App. \ref{app:dl}, squares of inner-dot products $b_{\pm2}\left(\vec{n},\vec{x}_i\right)$ are conveniently expressed as
\begin{equation}
	b_{\pm2}\left(\vec{n},\vec{x}_i\right)=\ds\sum_{\mu=1}^5b_\mu(\vec{x}_i)\,e^{(\pm)}_\mu(\vec{n}),
\end{equation}
where the 5 coefficients $b_\mu$ depends on the orientations of the magnetic fields only (i.e. they are functions of $\beta(\vec{x}_i)$ and $\alpha(\vec{x}_i)$ only). They are given by
\begin{equation}
	b_\mu(\vec{x}_i)=\left\{\begin{array}{l}
		\sqrt{\frac{2}{3}}(2\cos(\beta(\vec{x}_i))^2-\sin(\beta(\vec{x}_i))^2),\\
		-2\sin(\beta(\vec{x}_i))\cos(\beta(\vec{x}_i))e^{-i\alpha(\vec{x}_i)}, \\
		2\sin(\beta(\vec{x}_i))\cos(\beta(\vec{x}_i))e^{i\alpha(\vec{x}_i)},\\
		\sin(\beta(\vec{x}_i))^2e^{-2i\alpha(\vec{x}_i)},\\
		\sin(\beta(\vec{x}_i))^2e^{2i\alpha(\vec{x}_i)}.
	\end{array}\right.
\end{equation}
The 5 coefficients $e^{(\pm)}_\mu(\vec{n})$ are functions of the line-of-sight only and with our choice of the reference frame, they are expressed using spin-weighted spherical harmonics for $s=\pm2$ and $\ell=2$:
\begin{equation}
	e^{(\pm)}_\mu(\vec{n})=\sqrt{\frac{4\pi}{5}}\,\left\{\begin{array}{l}
		_{\pm 2}Y_{2,0}(\vec{\hat{n}}), \\
		_{\pm 2}Y_{2,1}(\vec{\hat{n}}), \\
		_{\pm 2}Y_{2,-1}(\vec{\hat{n}}), \\
		_{\pm 2}Y_{2,2}(\vec{\hat{n}}), \\
		_{\pm 2}Y_{2,-2}(\vec{\hat{n}}).
	\end{array}\right.
\end{equation}
Ensemble averages are done for the $b_\mu$ coefficients which for a uniform distribution of orientations gives $\left<b_\mu(\vec{x}_i)\,b_\nu(\vec{x}_j)\right>=\delta_{\mu,\nu}\,\delta_{i,j}$. The different correlation functions are then given by
\begin{eqnarray}
	\left<b_{\pm2}(\vec{n}_1,\vec{x}_i)b_{\pm2}(\vec{n}_2,\vec{x}_j)\right>&=&\frac{32\pi}{75}\delta_{i,j}\ds\sum_{m=-2}^2{}_{\pm2}Y_{2,m}(\vec{n}_1)\,{}_{\pm2}Y^\star_{2,m}(\vec{n}_2), \\
	\left<b_{\pm2}(\vec{n}_1,\vec{x}_i)b_{\mp2}(\vec{n}_2,\vec{x}_j)\right>&=&\frac{32\pi}{75}\delta_{i,j}\ds\sum_{m=-2}^2{}_{\pm2}Y_{2,m}(\vec{n}_1)\,{}_{\mp2}Y^\star_{2,m}(\vec{n}_2).
\end{eqnarray}
All these correlation are thus described by the angular power spectrum (which is nonzero for the 1-halo term only) reading $D^\perp_\ell=(32\pi/75)\,\delta_{\ell,2}$.

\end{document}